\documentclass[fleqn,usenatbib]{mnras}
\usepackage{newtxtext,newtxmath}
\usepackage[T1]{fontenc}
\DeclareRobustCommand{\VAN}[3]{#2}
\let\VANthebibliography\thebibliography
\def\thebibliography{\DeclareRobustCommand{\VAN}[3]{##3}\VANthebibliography}

\usepackage{graphicx}	
\usepackage{amsmath}	
\usepackage{xcolor}
\usepackage{tablefootnote}
\usepackage{orcidlink}
\usepackage{placeins}
\usepackage{caption}

\title[UV-IR SEDs of HRQs]{The rest ultraviolet to infrared spectral energy distributions of heavily reddened quasars are ``V-shaped'' and hot-dust poor}

\author[M. Stepney et al.]{Matthew Stepney,$^{\orcidlink{0000-0002-7711-0537}\,}$$^{1}$\thanks{E-mail: mstepney@cata.cl} Manda Banerji,$^{\orcidlink{0000-0002-0639-5141}\,}$$^{1}$ Shenli Tang$^{\orcidlink{0000-0002-2185-5679}\,}$$^{1}$ Matthew J. Temple, $^{\orcidlink{0000-0001-8433-550X}\,}$$^{2}$  and 
Paul C. Hewett$^{\orcidlink{0000-0002-6528-1937}\,}$$^{3}$  
\\
    $^{1}$School of Physics and Astronomy, University of Southampton, Southampton, SO17 1BJ, UK\\
    $^{2}$Centre for Extragalactic Astronomy, Department of Physics, Durham University, South Road, Durham DH1 3LE, UK\\ 
    $^{2}$Institute of Astronomy, University of Cambridge, Madingley Road, Cambridge CB3 0HA, UK\\
}

\date{Accepted XXXX. Received XXXX ; in original form XXXX}

\pubyear{2025}

\begin{document}
\label{firstpage}
\pagerange{\pageref{firstpage}--\pageref{lastpage}}
\maketitle

\begin{abstract}

\noindent We present a rest-ultraviolet to infrared spectral energy distribution (SED) analysis of 63 heavily reddened quasars (HRQs) at redshifts $0.7<\rm z_{sys}<2.7$ and with dust extinctions $0.4<\rm E(B-V)<1.8$. Our analysis demonstrates that SEDs with red optical and blue UV continua are very common in HRQs, with $>$82 per cent of the sample showing a UV-excess relative to the reddened quasar continuum. We model the SEDs by combining a reddened quasar and an unobscured scattered light component, though contributions from a star-forming host galaxy cannot be ruled out. The average scattering fraction is small ($\sim$0.3 per cent). Higher scattering fractions are ruled out by the $(i-K)>2.5$ colour-cut used to select HRQs which pre-dates the discovery of the \emph{JWST} “Little Red Dot” (LRD) population. Hence, LRDs generally have bluer UV continua. Nevertheless, four HRQs satisfy the LRD UV/optical continuum slope selections and are therefore massive, cosmic noon analogues of LRDs. Analysis of the near-infrared SEDs of HRQs reveals a deficit of hot dust relative to blue quasars, similar to what is observed in LRDs. This suggests HRQs trace a phase where strong AGN feedback processes eject dust from the inner torus. The UV scattering fraction of HRQs is weakly correlated with the amount of hot dust emission and anti-correlated with the line-of-sight extinction, $\rm E(B-V)$. This is consistent with the hot dust acting as the scattering medium, and the line-of-sight extinction being dominated by dust on interstellar medium scales in the host galaxy.

\end{abstract}

\begin{keywords}
quasars: general  --  galaxies: evolution  --  galaxies: photometry
\end{keywords}

\section{Introduction}

Dust-obscured accretion on to the central supermassive black holes in galaxies is thought to represent an important phase in the co-evolution of galaxies and their black holes \citep{1988ApJ...325...74S,Hopkins_2008,2005Natur.433..604D}. Over the last decade, the leap in sensitivity of wide-field optical and infrared surveys has led to the discovery of many different populations of dusty, obscured and red active galactic nuclei (AGN) in the high-redshift Universe. These include e.g. rQSOs \citep{Glikman:18, 2019MNRAS.488.3109K, Fawcett:22}, Extremely Red Quasars (ERQs; \citealt{2017MNRAS.464.3431H,Zakamska:16,2024MNRAS.527..950G}), Heavily Reddened Quasars (HRQs; \citealt{2012MNRAS.427.2275B,2015MNRAS.447.3368B,2019MNRAS.487.2594T}) and Hot Dust-Obscured Galaxies (Hot DOGs; \citealt{Eisenhardt_2012,2015ApJ...804...27A,2020ApJ...897..112A}) identified based on their colours in the observed optical, near- and mid-infrared wavelengths. Many of these luminous AGN would not be detected in the traditional ultraviolet (UV) and optical selections employed in previous quasar surveys like the Sloan Digital Sky Survey (SDSS), which is only sensitive to quasars with $\rm E(B-V)\lesssim 0.5$ \citep{Richards:03,2020MNRAS.494.4802F}, although deeper optical spectroscopic surveys e.g. with the Dark Energy Spectroscopic Instrument (DESI) are starting to push to higher extinctions \citep{2023MNRAS.525.5575F}. Together, these obscured and reddened quasars could account for a significant proportion of the black hole accretion budget \citep{2015MNRAS.447.3368B,2015ApJ...804...27A}. 

More recently, at even higher redshifts (e.g. $\rm z_{sys}\simeq4-8$), \textit{James Webb Space Telescope (JWST)} observations have also uncovered a new population of Little Red Dots (LRDs) selected based on their red rest-frame optical continuum colours and compact morphologies \citep{Kocevski_2023,2023ApJ...952..142F,2024Natur.628...57F,Labbe:25,Hviding:25}, with extremely high number densities - exceeding that of quasars by orders of magnitude \citep{2024arXiv240403576K,2024arXiv240610341A,2025arXiv250905434G}. While the true nature of some sub-samples of LRDs is still debated \citep[e.g.][]{2024ApJ...963..129M}, a significant proportion of LRDs show evidence for broad Balmer emission lines (e.g. FWHM $>$ 1000 $\rm  km\,s^{-1}$; \citealt{Kocevski_2023,Labbe:25,2025ApJ...989L...7T}) and are therefore likely dust-reddened AGN \citep{2024ApJ...964...39G, Wang25:RUBIES}. A puzzling feature of the LRD population are their ``V-shaped'' spectral energy distributions (SEDs), which display an upturn at rest-UV wavelengths, inconsistent with the red optical continuum (e.g. \citealt{Setton:24, 2024arXiv240403576K}). The V-shaped SEDs of LRDs have previously been modelled using two components e.g. a dust-reddened AGN superimposed on a blue component arising from a combination of scattered AGN light and/or an unobscured star-forming host galaxy \citep{2024ApJ...964...39G, Wang25:RUBIES}. However, recent observations point out that the inflection in the LRD SED is often associated with the Balmer limit \citep{Setton:24}. Together with the detection of Balmer absorption features in some LRD spectra \citep[e.g.][]{Ji:25b,D'Eugenio:25}, this suggests that the AGN may be cocooned by very dense, neutral, low-metallicity gas that is relatively dust-free \citep{2025ApJ...980L..27I,2025arXiv250316600D,2025arXiv250316596N}. This is also consistent with the exponential emission line profiles observed in LRDs \citep[e.g;][]{2025arXiv250316595R,2025arXiv251000103T,2025arXiv251107515K}, which suggest that the line broadening originates from electron scattering inside extremely compact regions with high electron column densities, as well as the apparent lack of hot and cold dust detections in many LRDs \citep{Setton:25}.

Since the discovery of the LRD population, searches for their lower redshift analogues have gathered pace \citep{Ji:25,Lin:25,Rinaldi:25,Ma:25}. However, many of the earlier populations of dusty and red AGN identified at cosmic noon and beyond in the pre-\textit{JWST} era also show similarities with the LRD SEDs. For example, $\sim$27 per cent of spectroscopically confirmed Hot DOGs display a blue excess \citep{Bao:25} with polarisation studies confirming  that scattered AGN emission is at least partially responsible for their blue rest-UV continua \citep{2022ApJ...934..101A}. Similarly, \citet{2018MNRAS.475.3682W} analysed a sample of 17 HRQs and demonstrated that at least 10 of them shown an upturn in the UV SEDs leading to a red rest-optical and blue rest-UV continuum. In \citet{2024MNRAS.533.2948S} we presented a deep 9hr XShooter spectrum of a prototype HRQ at $z\sim2.5$ - ULASJ2315 (J2315 hereafter) also known as `The Big Red Dot', which again shows an upward inflection in the UV superimposed on a red optical continuum. Broad UV emission lines detected in the X-Shooter spectrum suggest that scattered light from the obscured quasar contributes to the UV excess. However, additional constraints from the UV morphology and the rest optical spectrum indicate that unobscured regions in the star-forming host galaxy of the quasar also likely contribute to the UV light \citep{2024MNRAS.533.2948S}.

The SED of The Big Red Dot, J2315 has other similarities with the LRD population - e.g. weaker emission from hot dust, which could be interpreted as evidence that it is in a transitional phase between a post-merger starburst and an unobscured quasar when the dusty torus has been ``blown out" \citep{2024MNRAS.533.2948S}. In our standard evolutionary frameworks for massive galaxy evolution, earlier phases following a galaxy merger likely feature a reservoir of dense gas and dust surrounding the central engine. The GNz7q quasar may represent this early phase of galaxy evolution at the Epoch of Reionisation ($\rm z_{sys}\simeq7.2$), hosting compact, red rest-UV emission in addition to an extreme far-infrared continuum \citep{2022Natur.604..261F}. Furthermore, numerous populations at cosmic noon are consistent with this early evolutionary phase - e.g.  Hot DOGs \citep{2015ApJ...804...27A} and ERQs \citep{Goulding_2018} which both host very high gas column densities ($\rm N_H \sim 10^{23-24} \,cm^{-2}$). HRQs like J2315 could therefore correspond to an intermediate stage following this early cocooned phase, with more modest column densities of $\rm N_H \sim 10^{22} \,cm^{-2}$ \citep{2020MNRAS.495.2652L} and a higher proportion showing leaked or scattered emission in the rest-frame UV. 

It is clear that there is currently a huge diversity in red and obscured AGN populations with selection techniques evolving as new instruments and facilities become available. While the discovery of the \textit{JWST} LRD population has renewed interest in obscured AGN activity, many of the populations selected prior to \textit{JWST} could conceivably be similar to the LRD population \citep{2024MNRAS.533.2948S,Bao:25}. Exploring the similarities and differences in the multi-wavelength SEDs of the different samples of obscured AGN offers a route to better understanding the connection between these different populations. 

In this paper we focus on constructing and analysing the multi-wavelength rest-frame UV to infrared SEDs of a complete sample of 63 spectroscopically confirmed HRQs from \citet{2012MNRAS.427.2275B,2013MNRAS.429L..55B,2015MNRAS.447.3368B} and \citet{2019MNRAS.487.2594T} at $0.7<z<2.7$. We build on the work in \citet{2018MNRAS.475.3682W} and \citet{2024MNRAS.533.2948S} where a UV excess has been detected in some HRQs. The photometric data used in those works has now been superseded by deeper, more extensive observations probing the rest-frame UV SEDs of the majority of HRQs, which enables us to quantify the fraction that have a UV upturn similar to what has been observed in LRDs. We also fold in photometry from \textit{WISE} to constrain the rest-frame near infrared (NIR) SEDs of HRQs. The full and consistent SED analysis allows us to place HRQs in context with other obscured AGN populations such as Hot DOGs and LRDs, enabling the SED properties of dusty AGN to be explored over a large dynamic range in luminosity, obscuration level, black-hole mass and redshift.

The structure of this paper is as follows: Section \ref{Sec:data} outlines the core HRQ selection criteria and the assembly of new photometric data to construct their multi-wavelength SEDs. In Section \ref{sec:SED_Fits} we conduct the spectral energy distribution modelling of 63 HRQs before discussing our results in Section \ref{sec:C5_results}. Section \ref{sec:HRQ_DISC} describes the full ultraviolet to near-infrared SED properties of HRQs in the context of \emph{JWST}'s LRD population, as well as Hot DOGs, ERQs and blue quasars. Our main conclusions are summarised in Section \ref{SEC:HRQ_CONC}.

\section{Heavily Reddened Quasars (HRQs)} \label{Sec:data}

\subsection{Initial Selection}
\label{sec:selection}

Our parent population analysed in this paper consists of the spectroscopically confirmed HRQs from \citet{2012MNRAS.427.2275B}, \citet{2013MNRAS.429L..55B}, \citet{2015MNRAS.447.3368B} and \citet{2019MNRAS.487.2594T}. Their selection, based on observed near- and mid-infrared colours, is discussed in detail in these papers but briefly summarised here. HRQs pre-date the \textit{JWST} LRD population but are selected to have some very similar properties to LRDs in terms of their red rest-frame optical continuum colours and compact morphologies. The core selection utilises photometry from the United Kingdom Infrared Deep Sky Survey (UKIDSS), the Visible and Infrared Survey Telescope for Astronomy (VISTA) and the Wide-field Infrared Sky Explorer (\textit{WISE}). The selection criteria are as follows;

\begin{itemize}
    \item $K_{\rm AB} < 18.9\;\;|\;\;K_{\rm AB} < 20.3$\footnote{For the shallower surveys - e.g. the VISTA Hemisphere Survey \citep[VHS;][]{2013Msngr.154...35M} and the UKIDSS Large Area Survey \citep[UKIDSS-LAS;][]{2007MNRAS.379.1599L} - $K_{\rm AB} < 18.9$ mag was used. For the deeper VISTA Kilo-degree Infrared Galaxy survey \citep[VIKING;][]{2013Msngr.154...32E}, $K_{\rm AB} < 20.3$ mag was used.}
    \item $(J-K)_{\rm AB} > 1.6$
    \item $(W1-W2)_{\rm{Vega}}>0.85$
    \item $i_{\rm AB}>20.5$ (where available at the time from SDSS\footnote{Superseded in this paper by deeper optical photometry in many cases})
    \item $(i-K)_{\rm AB} > 2.5$ (where available at the time from SDSS)
    \item $kclass=-1$ (point-source morphology in the $K/K_S$-band)
\end{itemize}

Photometrically selected HRQ candidates from the infrared selections were followed up with observations on near-infrared spectrographs on Gemini and the Very Large Telescope \citep{2012MNRAS.427.2275B,2013MNRAS.429L..55B,2015MNRAS.447.3368B,2019MNRAS.487.2594T} leading to a spectroscopic confirmation of 63 HRQs with $0.7<z<2.7$ and $\rm E(B-V)\gtrsim 0.5$ that we analyse in this paper. These represent some of the most luminous and massive quasars known at cosmic noon, comparable to blue SDSS quasars, however, with levels of reddening consistent with \textit{JWST}'s less luminous LRD population. 

\subsection{Multi-wavelength Photometric Data}

In this section we describe the assembly of the photometric data needed to build the full multi-wavelength SEDs of all 63 HRQs. We make use of the near infrared photometry for all HRQs from UKIDSS and VISTA as well as their \textit{WISE} photometry in the \textit{WISE} $W1$ and $W2$ bands. Due to the point-source selection in the observed near-infrared, we utilise 2$\arcsec$ diameter aperture-corrected fluxes in the UKIDSS and VISTA bands, which are appropriate for point sources. To extend our photometric coverage to bluer wavelengths, we adopt a 1$\arcsec$ search radius to cross-match the HRQ sample with the following optical surveys: the Panoramic Survey Telescope and Rapid Response System \citep[Pan-STARRS - DR1;][]{2016arXiv161205560C}, the Sloan Digital Sky Survey \citep[SDSS - DR16;][]{2020ApJS..249....3A}, the Dark Energy Survey \citep[DES - DR2;][]{2021ApJS..255...20A}, HyperSuprime Camera Subaru Strategic Program \citep[HSCSSP - DR3;][]{2022PASJ...74..247A} and the Kilo-Degree Survey \citep[KiDS - DR5;][]{2024A&A...686A.170W}. Combining these surveys enables the study of the $ugriz-YJHK-W1W2$ photometry, where available, with an approximate rest-frame wavelength coverage of 1000\AA -- 3$\mu m$ at $\rm z_{\rm sys} \sim 2$. Several HRQs have been shown to have extended morphologies in the rest-UV \citep{2018MNRAS.475.3682W}. We therefore use CModel magnitudes for HSC; MAG$\_$AUTO for DECam, Pan-STARRS and SDSS; and Gaussian aperture magnitudes for KiDS. 

There are a total of seven sources for which we have rest-UV/optical data from all four optical surveys; Pan-STARRS, SDSS, DECam and HSC. In cases where the HRQs have photometric coverage in the same band from multiple surveys, we calculate weighted means across the bands to combine the photometry and boost the signal-to-noise ratio where possible. The weights are defined by the inverse variance of each observation. The name and weight of each filter contributing to a given photometric band is then recorded so that the photometry can be self-consistently modelled when fitting the spectral energy distributions (SEDs - see Section \ref{sec:SED_Fits} for details). The uncertainties in a given band are calculated using the standard error on the weighted mean. 

We find that the average percentage difference in fluxes between the different surveys and the weighted mean calculated for each band is $\sim10$ per cent. This difference is consistent with the known variability observed in quasar continua at rest-UV/optical wavelengths \citep[e.g.][]{1994MNRAS.268..305H}. Hence, to account for this variability, we impose a 10 per cent floor on the photometric uncertainties in bands for which there is only photometric data available from a single survey. Finally, we discard any photometric bands whose signal-to-noise ratio, $\rm S/N<3$. 

\section{Spectral Energy Distribution Modelling}\label{sec:SED_Fits}

To model the full rest-UV to near-infrared SEDs of the HRQs we utilise the \textsc{qsogen}\footnote{\url{https://github.com/MJTemple/qsogen}} tool, a Python package that implements an empirically-motivated parametric model to simulate quasar colours, magnitudes and SEDs \citep{2021MNRAS.508..737T}. To conduct the SED fitting we adopt a Monte-Carlo approach, using the \textsc{emcee}\footnote{\url{https://github.com/dfm/emcee}} Python package - which explores the likelihood space via the affine-invariant ensemble sampler proposed by \citet{2010CAMCS...5...65G}. 

Throughout the paper, we compare our results to a sample of blue SDSS quasars with redshifts $1.5<z_{\rm sys}<3.0$ \citep{2020MNRAS.492.4553R,2021MNRAS.501.3061T,2023MNRAS.523..646T}. \textsc{qsogen} models the hot dust emission with a single blackbody spectrum. To constrain both the amplitude and the effective temperature of the hot dust blackbody, a minimum of two filters must probe the hot dust emission. For this reason, \citet{2021MNRAS.501.3061T} imposes a $z_{\rm sys}<2.0$ selection on the SDSS DR16Q sample to ensure that both the $W1$ and $W2$ filters constrain the hot dust properties. Given that most HRQs have redshifts $z_{\rm sys}>2.0$, we fix the effective temperature of the hot dust blackbody  $\rm T_{bb} = 1243.6\,K$, as published in \citet{2021MNRAS.508..737T}, corresponding to a peak wavelength of $\lambda_{\rm peak} \simeq 2.3\mu m$\footnote{The version of \textsc{qsogen} used in \citet{2021MNRAS.501.3061T} 
was calibrated using a luminous subsample of DR14Q, leading to differences in the red power-law continuum slope and hot dust temperature when compared to model parameters calibrated in \citet[][see their Appendix A]{2021MNRAS.508..737T}. We re-fit the blue quasar sample with the same \textsc{qsogen} set-up as the HRQs to allow a direct and fair comparison. Hereafter; any reference to \citet{2021MNRAS.501.3061T} refers to blue SDSS quasars re-fit with the published version of \textsc{qsogen}.}. Since \textsc{qsogen} is not designed to reproduce dust emission at colder temperatures, we limit our SED fits to $1000\mathring{\rm A}< \lambda< 3\mu m$ in the rest frame.

\subsection{Single component dust-reddened quasar SED model} \label{sec:Single}

Initially, we chose a dust-reddened quasar SED model with three free parameters to fit the entire sample. The free parameters are (i) the 3000\AA\, continuum luminosity of the quasar, log$_{10}\{\lambda$L$_{\lambda} (3000\mathring{A}) [\rm erg\;s^{-1}]\}$ - (ii) the dust extinction, $\rm E(B-V)$ and (iii) the ratio in luminosity between the hot dust blackbody and the tail of the quasar UV/optical continuum slope at 2$\mu m$, $ L_{Dust}/L_{Disk}|_{2\mu m}$, which can be used as a proxy for the amplitude of the hot dust emission \citep[e.g. Figure 1;][]{2021MNRAS.501.3061T}. Hence, we permit the \textsc{emcee} package to explore an N-dimensional Gaussian likelihood function, where N represents the number of free parameters in the fit (N=3), and apply uniform priors. 

We assume the quasar extinction law discussed in Section 2.6 of \citet{2021MNRAS.508..737T}, which is similar to those derived by \citet{2004MNRAS.348L..54C} and \citet{2010A&A...523A..85G}. In addition, \textsc{qsogen}'s \emph{$emline\_type$} parameter controls the equivalent widths of the key quasar emission lines - such as H$\alpha$ and \ion{C}{iv} - giving the code the flexibility to vary the emission line contributions to the SED \citep{2021MNRAS.508..737T}. The code is constructed such that the $emline\_type$ parameter scales linearly with the observed EWs. Hence, we use the H$\alpha$ EWs reported in \citet{2012MNRAS.427.2275B,2015MNRAS.447.3368B,2019MNRAS.487.2594T,2024MNRAS.533.2948S} to fix the $emline\_type$ parameter in our simulations and use the H$\alpha$ EW and $emline\_type$ measurements of J2315 from \citet{2024MNRAS.533.2948S} to calibrate the relation. We use the default value - $emline\_type=0$ - for those objects for which we do not have H$\alpha$ measurements, as there is no evidence to suggest that the line EWs of HRQs differ from the blue SDSS quasars on which \textsc{qsogen} was developed \citep{2019MNRAS.487.2594T}.

For several of the HRQs, the observed photometric data comprise the weighted means of multiple surveys with different filter transmission curves. To ensure consistency between the modelled and observed data, we calculate the model photometry for each individual filter and combine like bands using the same weights as adopted when constructing the observed photometric catalogue in Section \ref{Sec:data}. Should a single photometric band dominate the $\overline{\chi^2_\nu}$ - i.e. a single band constitutes $>50$ per cent of the model uncertainty - the photometry is culled and the fit is rerun\footnote{This affects a total of five HRQs. In four sources the UKIDSS $J$-band is removed due to the very faint nature of the HRQ continua. In the remaining HRQ, the SDSS $r$-band is removed, which samples the \ion{C}{iii}] emission line.}. 

Following the single dust-reddened quasar SED fits, we calculate an average reduced chi-squared $\overline{\chi^2_\nu}|_{\rm Med} = 13.7$, which reduces to $\overline{\chi^2_\nu}|_{\rm Med} = 2.1$ when we consider only the 12/63 HRQs whose wavelength coverage does not extend blue-ward of 4000\AA.\, This both suggests that the average HRQ SED is inconsistent with a simple dust-attenuated quasar SED with a standard quasar extinction law, and that the primary driver of the poor SED fits is the rest-frame UV photometry. We next explore both a multi-component SED model in Section \ref{sec:multi} and a different dust extinction law in Section \ref{sec:orion} as an alternative.

\subsection{Dust-reddened quasar + scattered light SED model} \label{sec:multi}

In previous studies of smaller samples of HRQs, an excess in the UV emission has been seen given the inferred extinction at rest-optical wavelengths \citep{2018MNRAS.475.3682W,2024MNRAS.533.2948S}. This motivates the exploration of a two component SED model, which can reproduce the blue rest-UV photometric colours seen in at least some of the HRQs.  
The source of the UV excess observed in some dust-obscured quasars at cosmic noon is reportedly a combination of scattered quasar emission and/or ongoing star formation in the quasar host \citep[e.g.][]{2018MNRAS.475.3682W,2020ApJ...897..112A,2022ApJ...934..101A,2024MNRAS.533.2948S} - with the "leaked" quasar emission scenario being ruled out in HRQs and Hot DOGs due to the contrived dust geometries required \citep[see respective discussion sections;][]{2020ApJ...897..112A,2024MNRAS.533.2948S}. In our previous study of the 'Big Red Dot' HRQ (J2315) we found that scattered quasar emission and star formation in the quasar host provide equally good fits to the rest-UV continuum of the observed X-Shooter spectrum \citep{2024MNRAS.533.2948S}. As broad rest-frame UV emission lines are detected in several HRQs (\citealt{2024MNRAS.533.2948S} \& in prep) we adopt a scattered AGN SED to model the UV photometry in addition to the dust-attenuated quasar, though there is almost certainly a contribution to the UV flux from a star-forming host-galaxy given that many HRQs also host extended morphologies in the rest-UV \citep{2018MNRAS.475.3682W}. We carry out this two-component SED fit for the 51/63 HRQs whose wavelength coverage extends blue-ward of 4000\AA\, in the rest frame. 

The two-component SED model includes an additional free parameter (i.e. $\Delta p=1$), which represents the fraction of the total quasar luminosity scattered at rest-UV wavelengths, $F_{\rm UV}$. Only those objects whose wavelength coverage extends across $\geq 5$ photometric bands are re-fit with the updated model to ensure that the model is fully constrained. We impose the following criteria for the UV excess to be considered statistically significant - (i) the scattered component must contribute at least 50 per cent of the flux to the blue-most photometric band and (ii) the two-component SED model must represent an improvement over the single-component SED model to $>99$ per cent confidence - i.e. $\Delta \chi^2> 6.63$ \citep[See Table 1;][]{1976ApJ...210..642A}. Objects with a \emph{confirmed} UV excess satisfy both criteria. The presence of a UV excess is considered \emph{inconclusive} for objects that fail to meet criterion (i) - due insufficiently blue wavelength coverage - but do meet criterion (ii). Otherwise, the scattered component is \emph{rejected}.

\subsection{Single component dust-reddened quasar with Orion dust extinction law} \label{sec:orion}

As an alternative to the two-component SED model described in Section \ref{sec:multi}, some SED models explain the observed UV excess as an intrinsic feature of the rest-UV/optical continuum. One such example invokes atypical dust properties, where an obscuring medium with a deficit of smaller dust grains is unable to efficiently scatter emission at bluer wavelengths \citep[e.g.][]{2025ApJ...980...36L}. A scattering medium with such properties can be found in the Orion nebula \citep{1991ApJ...374..580B}, and some previous studies have found that the Orion dust extinction law can reproduce the stacked photometric data of LRDs without the need for an additional scattered/stellar emission component \citep{2025ApJ...980...36L}. Since \textsc{qsogen} has the flexibility to include a custom dust extinction law \citep{2021MNRAS.508..737T}, we also test whether the dust-reddened quasar SED model with the Orion dust extinction law used by \citet{2025ApJ...980...36L} can be used to fit the HRQ SEDs. We discuss these results in Section \ref{sec:UVExcess}. 
 
\subsection{Host galaxy contribution} \label{sec:galaxy_longlam}

Old stellar populations in the quasar host galaxy can make a significant contribution to the SED at $\lambda_{\rm rest}\sim 1\mu m$. The default host galaxy model used in \textsc{qsogen} is the S0 template from the SWIRE library \citep{2007ApJ...663...81P}, which provides a good fit to the average near infrared colours of blue SDSS quasars \citep[section 2.5;][]{2021MNRAS.508..737T}. Given that the S0 template was chosen to optimise the fit in the near-infrared, we do not expect it to accurately reproduce any quasar host galaxy contributions at rest-UV wavelengths, where the quasar accretion disk dominates in the blue quasars upon which \textsc{qsogen} was developed. Nevertheless, we test whether the rest-optical photometry of the HRQs require an additional component from old stellar populations in the quasar host. To determine the optimal galaxy fraction ($fragal$), we model the SEDs of the HRQ sample with the following values; $fragal=0.01, 0.03, 0.05, 0.07, 0.09$. We then calculate the average reduced chi-squared statistic for each model to estimate the optimum host galaxy fraction for the HRQ sample.

\section{Results} \label{sec:C5_results}

We find that 51 of the 63 HRQs have photometry blue-ward of $\sim$4000\AA\ where a two-component SED model can be tested. The remaining 12 are fit with the single component dust-reddened quasar SED model described in Section \ref{sec:Single}. The final best-fit parameters for all 63 HRQs are presented in Appendix \ref{App:plots}. The median reduced chi squared statistic of all 63 HRQ SED fits is $\overline{\chi^2_\nu}|_{\rm Med} = 1.9$.

\subsection{Luminosities, dust extinctions \& Eddington ratios}
\label{sec:properties}

While previous studies of HRQs \citep{2012MNRAS.427.2275B,2013MNRAS.429L..55B,2015MNRAS.447.3368B,2019MNRAS.487.2594T} have presented estimates for their line-of-sight dust extinction, $\rm E(B-V)$ and implied dust-corrected luminosities, these results are superseded by the current work which incorporates more extensive photometry in almost all cases and provides self-consistent fits from the rest-UV to the infrared. In general, the addition of a second scattered AGN component to explain the upturn in the SEDs at rest-UV wavelengths leads to higher values for the dust extinction of the attenuated component relative to the previous works, which did not detect and model the UV excess emission in HRQs. This in turn leads to higher inferred dust corrected luminosities for the HRQs. 

\begin{figure}
\centering
 \includegraphics[scale=0.63, trim={4.0cm, 0cm, 4.0cm, 0.7cm}]{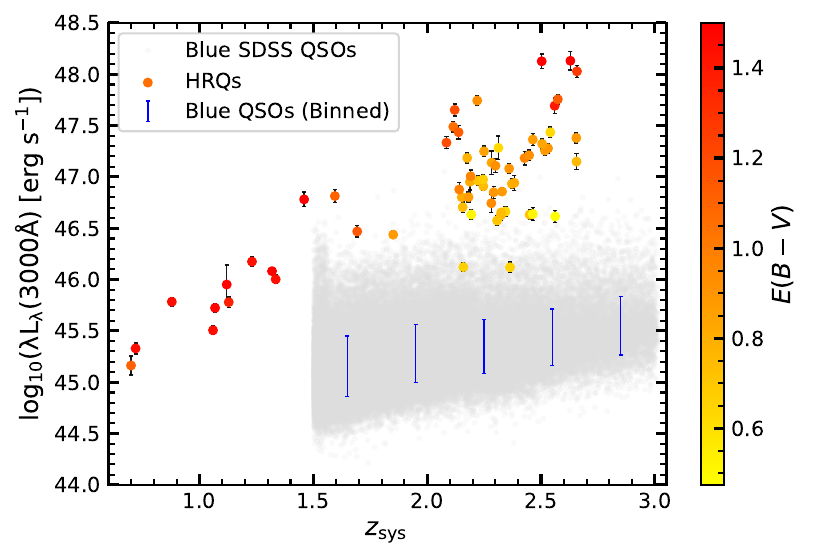} 
  \caption{The redshift versus 3000\AA\, continuum luminosity for blue SDSS quasars \citep[grey;][]{2020MNRAS.492.4553R,2021MNRAS.501.3061T,2023MNRAS.523..646T} and HRQs (coloured circles). The colour bar illustrates the $\rm E(B-V)$ of each HRQ. }
 \label{fig:L3000_Zsys}
\end{figure}

\autoref{fig:L3000_Zsys} shows the redshift versus dust-corrected 3000\AA\, continuum luminosities of the HRQ sample compared to blue SDSS quasars whose SEDs were fit with a consistent methodology \citep{2020MNRAS.492.4553R,2021MNRAS.501.3061T,2023MNRAS.523..646T}. HRQs are more luminous than their blue SDSS counterparts at redshifts $\rm z_{sys}>1.5$ after correcting for the dust extinction due to the flux limits of the surveys. Furthermore, the lower redshift HRQs are more heavily obscured than their higher redshift counterparts because of the near-infrared HRQ colour-selection. 

\begin{figure}
\centering
 \includegraphics[scale=0.62, trim={4.0cm, 0.0cm, 4.0cm, 0.5cm}]{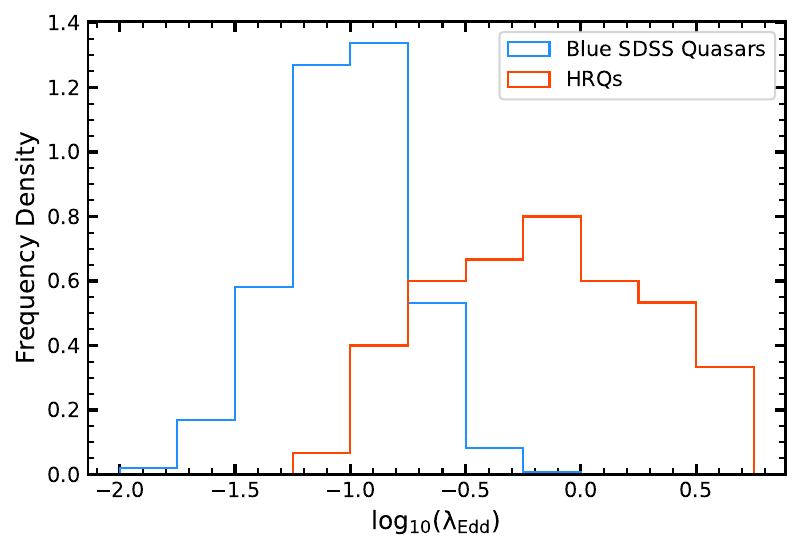} 
  \caption{The Eddington-scaled accretion rate distributions of the HRQ sample (orange) and blue SDSS quasars \citep[blue;][]{2020MNRAS.492.4553R,2023MNRAS.523..646T}. HRQs generally host higher accretion rates than their blue SDSS counterparts, with HRQs tending towards the super-Eddington regime.}
 \label{fig:LEdd}
\end{figure}

Using previously reported estimates of the black hole mass \citep{2012MNRAS.427.2275B,2013MNRAS.429L..55B,2015MNRAS.447.3368B,2019MNRAS.487.2594T,2024MNRAS.533.2948S} combined with the updated luminosities calculated in this work, we determine the Eddington-scaled accretion rates for the entire HRQ sample, using Eqn. \ref{eq:kbol} to estimate a bolometric correction for each individual HRQ. 

\begin{equation}
    k_{\rm Bol} = 25\{\lambda L_\lambda(3000\mathring{\rm A})/10^{42}\,\rm [erg\,s^{-1}]\}^{-0.2}
    \label{eq:kbol}
\end{equation}

\noindent where $\rm \lambda L_\lambda(3000$\AA) is the extinction-corrected 3000\AA\, continuum luminosity of the HRQ \citep{2019MNRAS.488.5185N}. \autoref{fig:LEdd} illustrates that HRQs extend into the super-Eddington regime, with accretion rates an order of magnitude higher than their blue SDSS counterparts at similar redshift\footnote{We also use the \citet{2019MNRAS.488.5185N} relation to determine $k_{\rm Bol}$ and therefore updated luminosities and Eddington ratios for all SDSS quasars. The black hole masses of the blue SDSS quasars were derived using \ion{Mg}{ii}, rather than Balmer lines for the HRQ sample. As both mass estimators are broadly consistent \citep[e.g.][]{2012ApJ...753..125S}, the difference in mass estimator cannot explain the result presented in Fig. \ref{fig:LEdd}.}.

\subsection{Ubiquitous excess UV emission in HRQs} \label{sec:UVExcess}

\begin{figure}
\centering
 \includegraphics[scale=0.355, trim={4.0cm, 0.6cm, 4.0cm, 0.2cm}]{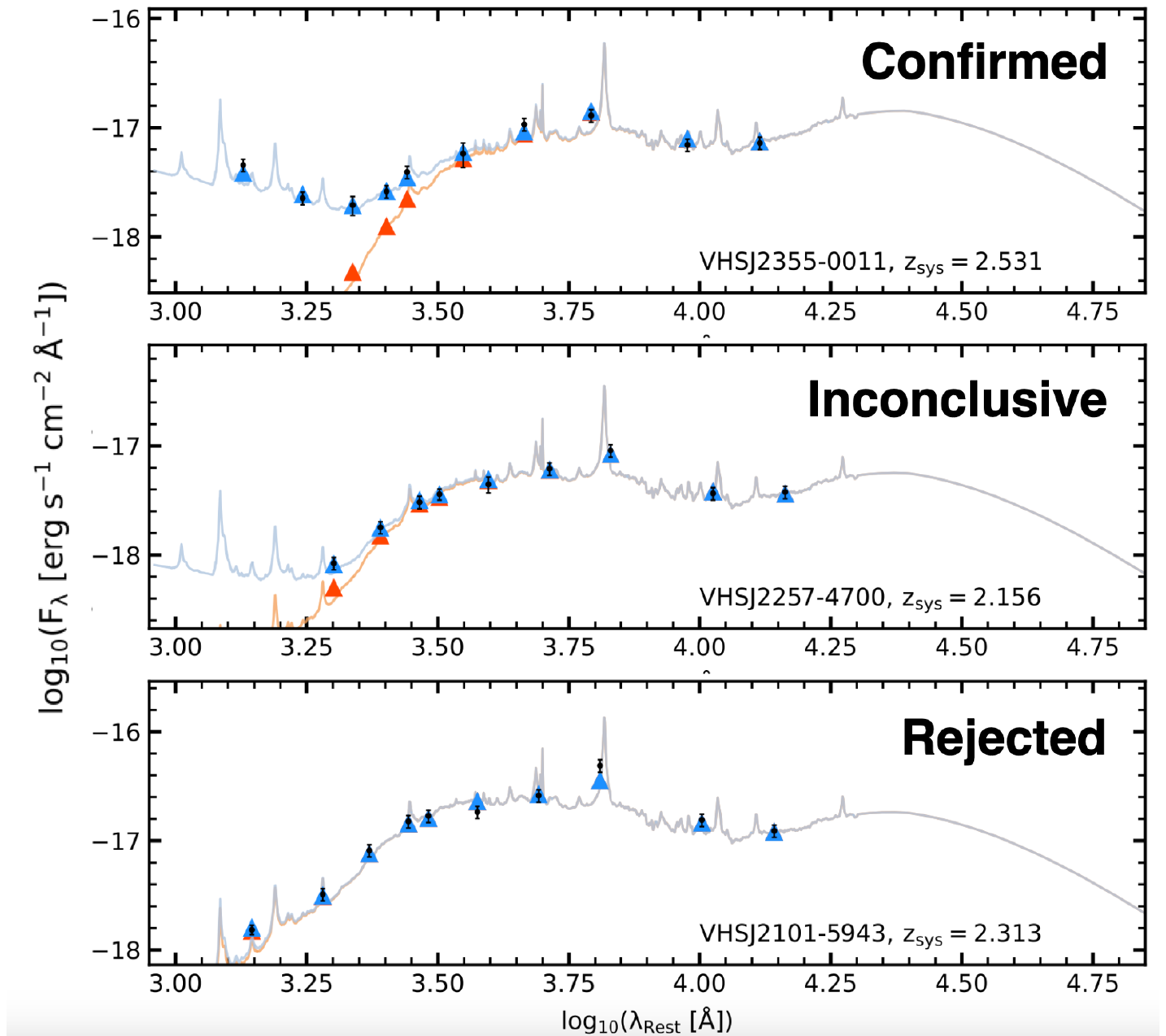} 
  \caption{Example SED fits for HRQs where a UV excess is "confirmed" (top), "inconclusive" (middle) or "rejected" (bottom). The photometric data and associated uncertainties are presented in black. The best-fit SED model and photometry are presented by the blue line and triangles. The dust attenuated quasar component and photometry are presented by the orange line and triangles. In the "confirmed" case, there is clear evidence of a UV excess. In the "inconclusive" case, there is tentative evidence of a UV excess, but the wavelength coverage does not extend blue enough to constrain the model. In the "rejected" case, there is no evidence to support that the two-component model yields a better fit than a single-component reddened quasar SED.}
 \label{fig:SEDs_UVExcess}
\end{figure}

Based on the SED fits, a statistically significant UV excess is confirmed in 42/51 of the HRQs for which photometric data blueward of 4000\AA\, was available, representing 82 per cent of the sample. This is a much larger fraction than seen in e.g. Hot DOGs where $<$30 per cent show a UV excess \citep{Bao:25}. Objects for which a statistically significant UV excess was rejected represent just 4/51 of the sample, or equivalently 8 per cent. The remaining 5/51, or equivalently 10 per cent, of the HRQs are considered \emph{inconclusive} by our analysis. Typical examples of the \emph{confirmed}, \emph{inconclusive} and \emph{rejected} UV excess cases are represented in Fig. \ref{fig:SEDs_UVExcess}. In the remaining 12 HRQs we do not have photometric data blue enough to assess whether there is a UV excess. 

Our analysis suggests that excess UV emission leading to an upward inflection or ``V-shape" in the SEDs is therefore extremely common amongst HRQs despite the initial selection (Section \ref{sec:selection}) actively biasing against objects that are UV-bright. While this result does not confirm that scattered light is the primary source of the UV excess in HRQs as the scattered quasar light and the star-forming host galaxy interpretations of the UV excess are degenerate \citep{2018MNRAS.475.3682W,2020ApJ...897..112A,2024MNRAS.533.2948S}, it does demonstrate that the excess rest-UV emission can be effectively modelled using a two-component SED fit. 

\begin{figure}
\centering
 \includegraphics[scale=0.64, trim={4.0cm, 0.0cm, 4.0cm, 0.0cm}]{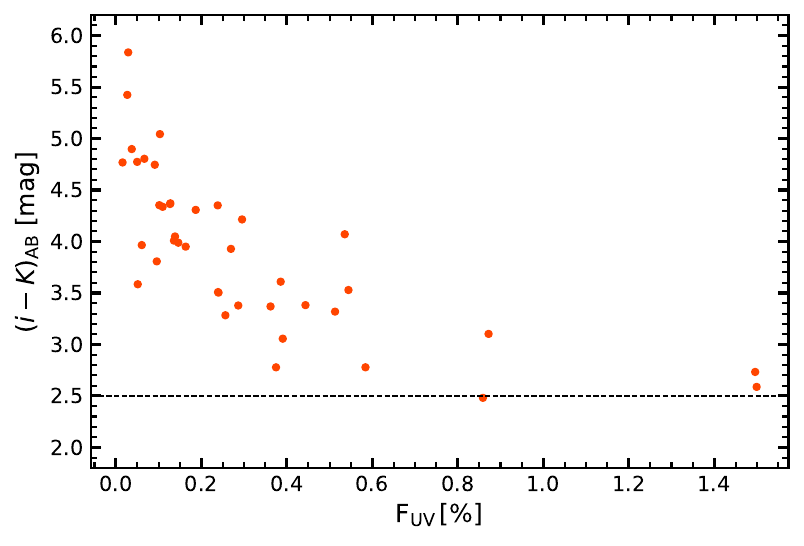} 
  \caption{The scattering fraction, $F_{\rm{UV}}$, vs the $(i-K)_{\rm AB}$ colour, calculated from the best-fit SED models for the 42 HRQs that exhibit a statistically significant UV excess. The $(i-K)_{\rm AB} =2.5\,\rm mag$ colour selection is represented by dashed black line. There is a strong anti-correlation between $F_{\rm{UV}}$ and the $(i-K)_{\rm AB}$ colour. Therefore, this colour selection excludes HRQs with higher scattering fractions from the sample.}
 \label{fig:i-band_selections}
\end{figure}

The average scattering fraction (introduced in Section \ref{sec:multi}) of the HRQ sample - $\langle F_{\rm{UV}}\rangle_{\rm HRQ} = 0.26$ per cent - is low when compared to other red AGN populations \citep[e.g. $\sim 3$ per cent in LRDs;][]{2024ApJ...964...39G}. Since the HRQ selection was derived prior to the discovery of blue photometric colours in their rest-UV continua \citep{2018MNRAS.475.3682W}, the $(i-K)_{\rm AB}$ colour selection removes sources with a higher scattering fraction, $F_{\rm{UV}}$, from the sample. \autoref{fig:i-band_selections} shows the strong anti-correlation between the scattering fraction, $F_{\rm{UV}}$, and the $(i-K)_{\rm AB}$ colour. We estimate that HRQs whose scattering fractions exceed $F_{\rm UV}\gtrsim 1$ per cent would fail the $(i-K)_{\rm AB}$ colour selection and are therefore removed from the sample. 

In Section \ref{sec:orion} we discuss an alternative description of the UV excess which invokes Orion-like dust properties to reproduce blue rest-UV continua. We find not a single HRQ favours the Orion dust extinction law over the dust-reddened quasar + scattered light SED model. A full discussion can be found in Appendix \ref{App:orion}.

\subsection{Contribution from old stellar populations at $z_{\rm sys}<1.5$} \label{sec:gal_IR}

We find that the inclusion of the S0 template in the SED fit offers a significant improvement over the standard dust-reddened quasar + scattered light SED model at $\rm z_{\rm sys}<1.5$ but not at higher redshifts. The addition of the host galaxy component with the best galaxy fraction, $fragal=0.05$, yields a median reduced chi-squared for the $z_{\rm sys}<1.5$ sub-sample $\overline{\chi^2_\nu}|_{\rm Med} = 3.0$, increasing to $\overline{\chi^2_\nu}|_{\rm Med}  = 6.8$ when the S0 template is omitted. In contrast, the SED fits of HRQs with redshifts $z_{\rm sys}>1.5$ are not improved by the inclusion of the S0 template. Assuming a galaxy fraction, $fragal=0.05$, yields an median reduced chi-squared $\overline{\chi^2_\nu}|_{\rm Med}  = 2.3$, decreasing to $\overline{\chi^2_\nu}|_{\rm Med}  = 1.9$ when the S0 template is omitted. For this reason, we only include a host galaxy component in the SED fits for those HRQs where $z_{\rm sys}<1.5$.

\begin{figure}
\centering
 \includegraphics[scale=0.31, trim={0.0cm, 1.6cm, 0.0cm, 1.7cm}, clip]{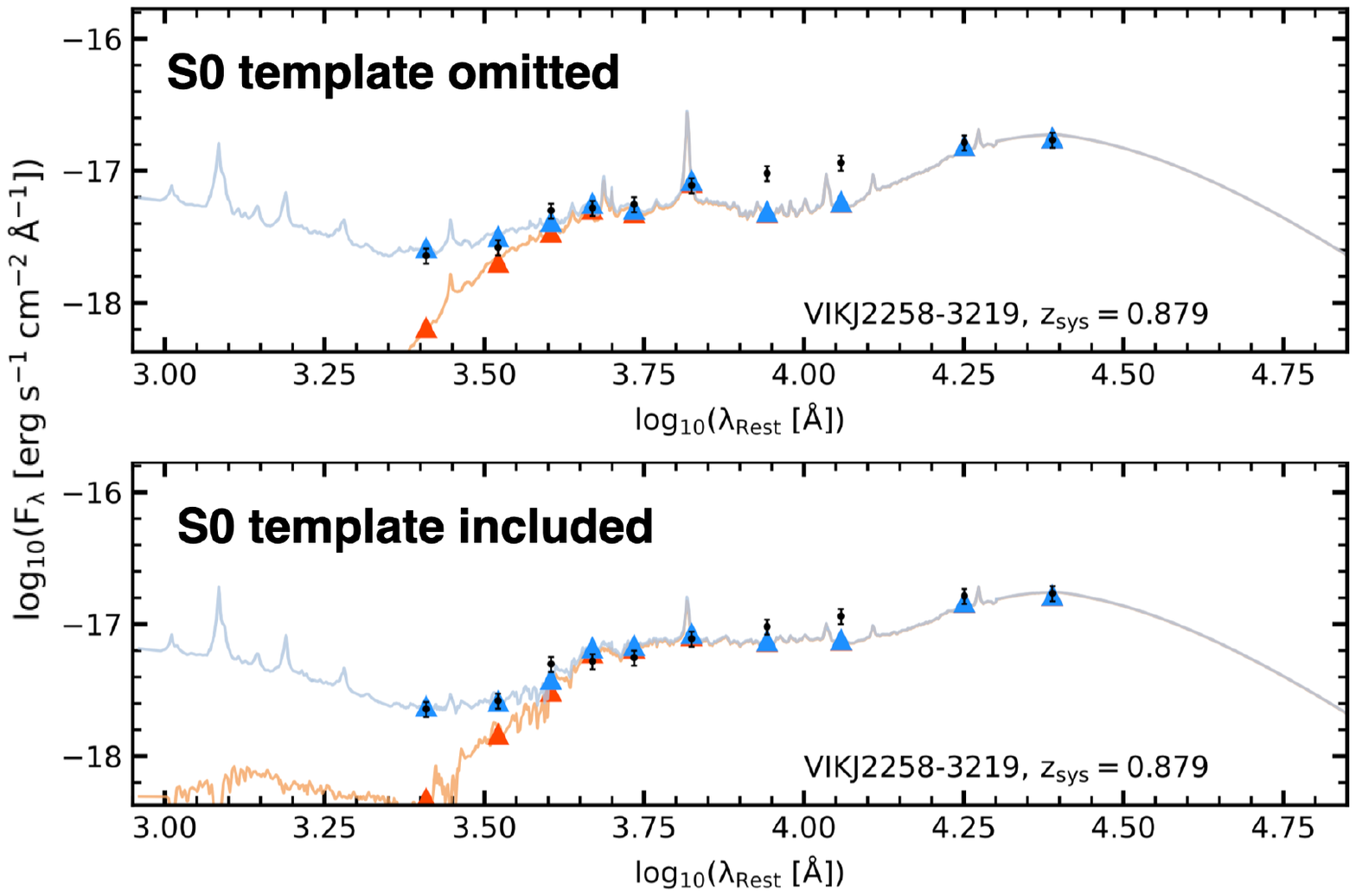} 
  \caption{The best-fit SED for the HRQ VIKJ2258-3219 at $z=0.879$ with the S0 galaxy template omitted (top) and with $fragal=0.05$ (bottom). The photometric data and associated uncertainties are presented in black. The best-fit SED model and photometry are indicated by the blue line and triangles. The dust-attenuated quasar component and photometry are indicated by the orange line and triangles. At near-infrared wavelengths the model is unable to reproduce the photometry in the top panel, suggesting that old stellar populations have a significant contribution to the SED in this region. Hence, the inclusion of the galaxy component improves the fit in the bottom panel.}
 \label{fig:lowz_HRQ}
\end{figure}

An example SED of a $z<1.5$ HRQ is presented in Fig. \ref{fig:lowz_HRQ}, illustrating how contributions from old stellar populations in the quasar host galaxy can significantly improve the SED model at infrared wavelengths. The dichotomy in the sample at $z_{\rm sys}\simeq1.5$ may result from the significant luminosity difference between the high- and low-redshift samples (e.g. Fig. \ref{fig:L3000_Zsys}). The average optical luminosity of the $z_{\rm sys}<1.5$ sample is log$_{10}\{\lambda$L$_{\lambda} (3000\mathring{A})\;[\rm erg\;s^{-1}]\} = 45.8$ compared to log$_{10}\{\lambda$L$_{\lambda} (3000\mathring{A})\; [\rm erg\;s^{-1}]\}$ $= 47.1\;$ at $z_{\rm sys}>1.5$ - meaning the quasar emission is less likely to dominate the host in the lower-redshift sub-sample. The lower-redshift sub-sample also probes redder HRQs with higher $\rm E(B-V)$ than those observed at $z_{\rm sys}>1.5$. The higher extinctions in the low-z sample may therefore reveal a greater fraction of the host galaxy emission if the host galaxy is less obscured than the AGN. In either case, the stellar continuum in the lower-redshift objects would be expected to contribute more significantly to the SED at near-infrared wavelengths as we find.

\subsection{A deficit of hot dust in HRQs} \label{sec:dust}

With \textit{WISE} photometry available for all sources in the HRQ sample, it is possible to constrain the hot dust emission of the entire population. The mean effective temperature of the hot dust blackbody in blue SDSS quasars (and therefore the temperature assumed for the HRQ SED fitting, $\rm T_{bb} \simeq 1243\,K$), is consistent with the sublimation-temperature of silicate dust \citep{article,2021MNRAS.501.3061T}, the maximum permitted dust temperature before the dust grains are destroyed by the quasar radiation field. Hence, the relative amplitude of the sublimation-temperature dust emission with respect to the tail of the accretion disk blackbody (i.e. $\rm L_{Dust}/L_{Disk}|_{2\mu m}$) is a good measure of the amount of dust present on 10s-of-pc scales \citep{2021MNRAS.501.3061T}.

In Appendix \ref{App:IR_Selection}, we discuss how the colour selection of HRQs described in Section \ref{sec:selection} affects the distribution of their sublimation-temperature dust amplitudes. We find that at redshifts $z_{\rm sys}\geq1.5$, $\rm L_{Dust}/L_{Disk}|_{2\mu m}$ is insensitive to the $(J-K)_{\rm AB}$ colour selection discussed in Section \ref{Sec:data}. Given that our treatment of the host galaxy contribution is also different at $z_{\rm sys}=1.5$, we consider only the 50 HRQs with redshifts $z_{\rm sys}\geq1.5$ and $\overline{\chi^2_\nu} < 5.0$ when discussing the hot dust properties of the HRQ sample. 

\begin{figure}
\centering
 \includegraphics[scale=0.63, trim={0cm, 0.35cm, 0cm, 0.1cm}, clip]{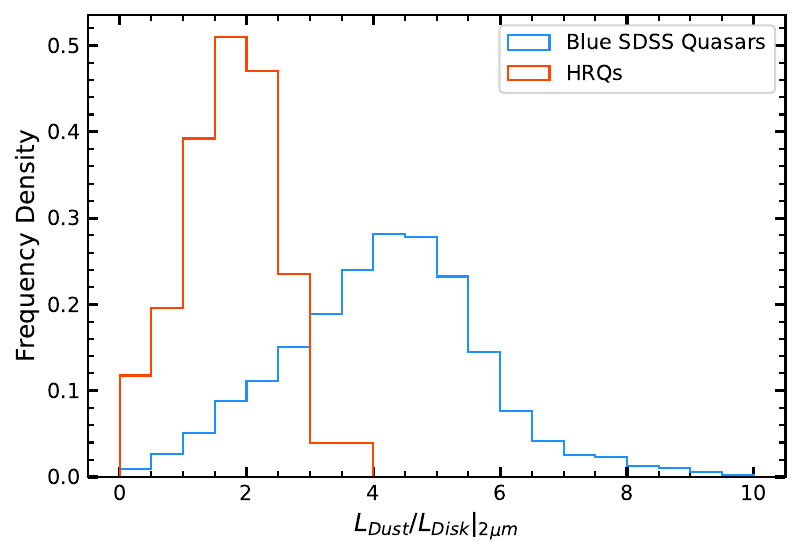} 
  \caption{Histograms illustrating the ratio in luminosity between the hot dust blackbody and the tail of the UV continuum slope at 2$\mu m$. The $z_{\rm sys}\geq1.5$ HRQ sample is shown in orange and the blue SDSS quasars studied in \citet{2021MNRAS.501.3061T} are shown in blue. The HRQ histogram is skewed toward lower sublimation-temperature dust amplitudes, implying that there is less dust in the inner regions of HRQs with respect to blue quasars.}
 \label{fig:Dust_amps}
\end{figure}

\autoref{fig:Dust_amps} illustrates the distribution of hot dust amplitudes for these 50 HRQs with redshifts $z_{\rm sys}\geq1.5$ and $\overline{\chi^2_\nu} < 5.0$ compared to the blue SDSS quasars studied by \citet{2021MNRAS.501.3061T}\footnote{As mentioned in Section \ref{sec:SED_Fits}, the blue SDSS quasars studied by \citet{2021MNRAS.501.3061T} are restricted in redshift to $z_{\rm sys}<2.0$. Given there is no evidence to suggest that the hot-dust properties evolve with redshift \citep[For a full discussion, see appendix D;][]{2021MNRAS.501.3061T}, this does not bias our results.}. The mean sublimation-temperature dust amplitude of the 50 HRQs with $z_{\rm sys}\geq1.5$ is $\langle \rm L_{Dust}/L_{Disk}|_{2\mu m} \rangle = 1.6\pm0.8$, lower than the $4.4\pm2.1$ observed in typical blue SDSS quasars \citep{2021MNRAS.501.3061T}. Furthermore, Fig. \ref{fig:Dust_amps} illustrates that HRQs have a significantly more modest dynamic range in $\rm L_{Dust}/L_{Disk}|_{2\mu m}$ when compared to blue SDSS quasars - with a 99$^{th}$ percentile of $\rm \left[L_{Dust}/L_{Disk}|_{2\mu m}\right]_{99^{th}}=3.5$ and $\rm \left[L_{Dust}/L_{Disk}|_{2\mu m}\right]_{99^{th}}=13.0$ for the HRQ and blue SDSS quasar samples, respectively. A two-sided Kolmogorov-Smirnov (KS) test yields a $p$-value = $2\times10^{-30}$, suggesting that the two distributions are statistically distinct. In Appendix \ref{App:IR_Selection} we discuss how the \mbox{$(W1-W2)$} colour selection may bias the HRQ sample to higher dust amplitudes - with HRQs at the average dust extinction requiring $\rm L_{Dust}/L_{Disk}|_{2\mu m}>1.0$ to be selected at $z_{\rm sys}\gtrsim2$. It is therefore likely that HRQs with the lowest hot dust amplitudes may have escaped our selection, serving only to amplify the difference between the distributions. This result implies that HRQs host significantly less hot dust compared to blue quasars, or that the emission from the hottest dust has been self-absorbed by cooler dust components as is the case in Type 2 quasars. These possibilities are discussed further in Section \ref{sec:blueQSOs}.

\subsection{A pan-chromatic composite SED for HRQs} \label{sec:stacked_photo}

\begin{figure*}
\centering
 \includegraphics[scale=1, trim={0.0cm, 0.2cm, 0.0cm, 0.2cm}, clip]{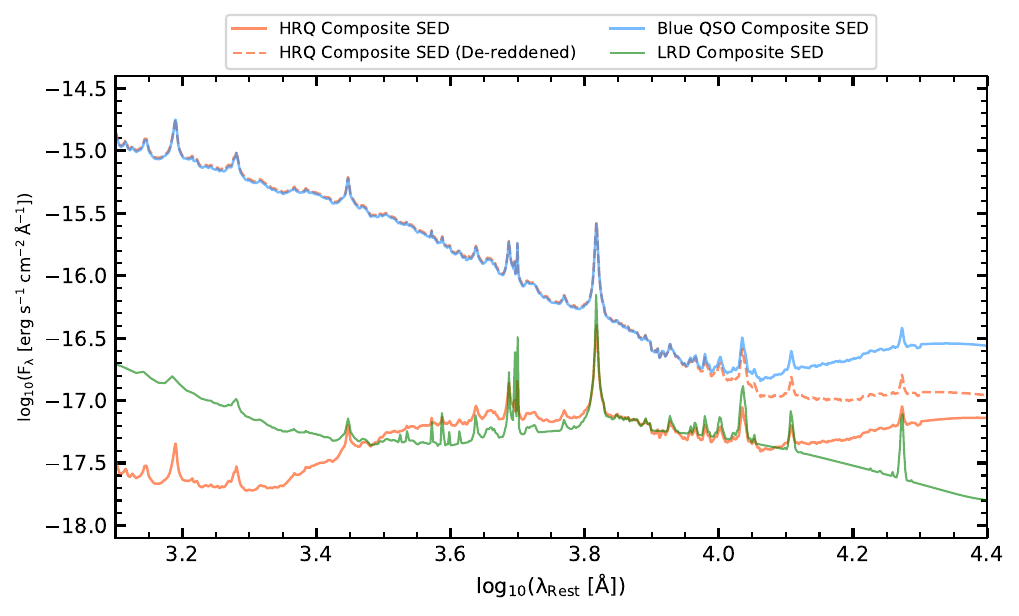} 
  \caption{We present composite SEDs for HRQs (orange), LRDs \citep[][green]{2024arXiv240610341A} and blue SDSS quasars \citep[][blue]{2021MNRAS.501.3061T} - with the de-reddened HRQ composite SED (orange, dashed) for reference. The LRD and HRQ composite SEDs are normalised to the median flux in the window $\rm log_{10}(\lambda_{Rest} \; [\mathring{A}]) = [3.88,3.92]$. The de-reddened HRQ and blue SDSS quasar SEDs are also normalised to the median flux in the window $\rm log_{10}(\lambda_{Rest} \; [\mathring{A}]) = [3.88,3.92]$. HRQs and LRDs have similar SED shapes in the rest optical and both populations feature a UV excess. Both HRQs and LRDs feature supressed hot dust emission with respect to luminous blue quasars.}
 \label{fig:Stacked_Phot_HRQ}
\end{figure*}

We construct a noise-weighted composite SED from the individual best-fit SEDs of HRQs in the rest frame. The composite includes only the HRQs with redshifts $z_{\rm sys}\geq1.5$, given that the near-infrared portion of the SED is handled differently in the lower redshift sample, and wavelength coverage blue enough to robustly test for a UV excess. The composite is therefore comprised of 42 individual HRQs. The inverse square of the $\overline{\chi^2_\nu}$ statistic of each SED fit is used as the weights for the stacking. The results are presented in Fig. \ref{fig:Stacked_Phot_HRQ} and will also be made available online as supplementary material. 

As expected from Sections \ref{sec:UVExcess} $\&$ \ref{sec:dust}, we observe a significant UV excess consistent with scattered quasar emission and a weak contribution from the hot dust. We compare our composite HRQ SED to a noise-weighted composite of the LRD population - produced using COSMOS photometry \citep{2024arXiv240610341A}. The composite SEDs show that the populations appear consistent at rest-optical wavelengths and both feature suppressed emission at $\sim 2\mu m$ with respect to blue SDSS quasars, although there is a stronger contribution from hot dust in the HRQs compared to LRDs. At rest-UV wavelengths, there is a clear difference in the average SED shapes of the HRQ and LRD populations - with LRDs featuring much bluer continuum slopes than the average HRQ. As discussed in Section \ref{sec:UVExcess}, this is due to the $(i-K)$ colour selection applied to the HRQ sample, which precludes HRQs with high scattering fractions and blue rest-UV continua. After correcting the HRQ SED for the line-of-sight extinction, $\rm E(B-V)$, the de-reddened HRQ composite SED clearly confirms our result from Section \ref{sec:dust}, whereby blue SDSS quasars \citep{2021MNRAS.501.3061T} feature a stronger hot dust contribution than HRQs. 

\section{Discussion} \label{sec:HRQ_DISC}

In this section, we discuss our results in the context of other AGN and quasar populations focusing in particular on obscured and reddened AGN in the high-redshift Universe such as Hot DOGs and Little Red Dots. Having constrained the multi-wavelength SED properties from the UV to the infrared, we also look at the links between dust extinction, scattering and dust emission in HRQs. Finally, we discuss how the HRQ selection criteria might need to be amended in light of the results from this paper, so a more complete sample of dust-reddened quasars can be identified at cosmic noon.  

\subsection{The dust emission properties of HRQs} \label{sec:blueQSOs}

In Section \ref{sec:dust} we find that HRQs exhibit systematically lower hot dust emission compared to blue SDSS quasars after correcting for the line-of-sight extinction inferred from their red optical continuum colours. This suggests that either the HRQ population has less dust at the sublimation radius, or the inner torus emission has been self-absorbed by even higher columns than inferred from the attenuation at rest-frame optical wavelengths. In the local Universe, interferometric observations have resolved the torus emission into two distinct components - an equatorial hot disk and a more extended warm polar wind \citep{Stalevski:17,Hoenig:17}. Radiative transfer modelling of the torus emission can also explain the diversity in the mid infrared SEDs of AGN in the local Universe with these two components of the dust emission \citep{Hoenig:19}. One possibility is that the low hot-dust amplitudes of HRQs provide evidence that they are observed in a ``blowout" phase when strong radiative feedback processes have depleted the hottest disk component of the torus leading to most of the dust existing on larger, galaxy-wide scales potentially in a warm polar wind component. This picture would also be consistent with the large far infrared luminosities of some HRQs \citep[e.g.][]{2014MNRAS.439L..51B,2018MNRAS.479.1154B} and the fact that HRQs generally accrete at Eddington/super-Eddington rates (Fig. \ref{fig:LEdd}). 

The other possibility is that the hottest component of the torus is self-absorbed as is often the case in Type 2 AGN \citep{Hoenig:19, Hickox:17}. We consider this possibility less likely for two reasons. The detection of very broad emission lines in the rest-frame optical spectra of all HRQs \citep{2012MNRAS.427.2275B,2015MNRAS.447.3368B,2019MNRAS.487.2594T} rules out completely edge-on sight lines. Independent constraints on the gas column densities of HRQs from X-ray spectroscopy also suggest that the obscuring columns are modest, and that HRQs are Compton-thin rather than Compton-thick AGN \citep{2020MNRAS.495.2652L} - consistent with the SED-inferred dust extinctions from this paper. Finally, the dust-corrected luminosities of HRQs are already very high (Section \ref{sec:properties}) and even larger dust corrections to these luminosities would lead to unphysical values in some cases.  

It is important to note that the mid-infrared colour selection of the HRQs, $(W1-W2) > 0.85$ (Vega) is redder than the typical colours of \textit{WISE} AGN e.g. the 90 per cent completeness cut in \citet{Assef18:WISE} is $(W1-W2) > 0.5$ (Vega). As discussed in detail in Appendix \ref{App:IR_Selection}, the \textit{WISE} colour cut preferentially selects quasars with higher hot dust amplitudes at $z_{\rm sys}>2$. The fact that we nevertheless find that the HRQs in our sample are hot-dust poor, indicates that there may be more dust-poor reddened quasars in the high-redshift Universe that have escaped the HRQ \textit{WISE} colour selection. 

\subsubsection{Hot dust depletion \& connection to AGN feedback}

Should a "blow-out" phase be driving the low hot dust amplitudes in HRQs, we might expect to observe similarly weak dust emission in other obscured AGN populations. For example, LRDs host flat mid-infrared SEDs with very little hot dust emission \citep{2024arXiv240610341A}, as can be seen in Fig \ref{fig:Stacked_Phot_HRQ}, and like HRQs, have more modest gas column densities with respect to Hot DOGs and ERQs \citep[e.g.][]{2025arXiv250316600D,2025ApJ...989L...7T,2025arXiv250316596N,2025arXiv251107515K}. However, LRDs are also weak far-infrared emitters and can host very low metallicities \citep{2024arXiv240610341A,2025arXiv250316600D,Setton:25, 2025arXiv250905434G}, suggesting that dust may be entirely absent from their obscuring media, with metal-poor gas primarily responsible for enshrouding their central engines \citep{2025arXiv250316596N}. On the other hand, while Hot DOGs host extremely red continua, with very high dust extinctions $\rm E(B-V) = 2-20$ \citep{2015ApJ...804...27A}, their selection is biased towards mid-infrared bright and therefore dust-rich sources by design \citep[e.g. $(W2-W3) > 5.3$ mag and $W3<10.6$ mag;][]{2015ApJ...804...27A}. Hence, they are probably dominated by AGN where the hottest dust is self-absorbed, rather than missing entirely, although there is some evidence that Hot DOGs with a blue UV excess also host extended polar, dusty outflows \citep{2025A&A...702A.124A}.  
 
\begin{figure}
\centering
\includegraphics[scale=0.6]{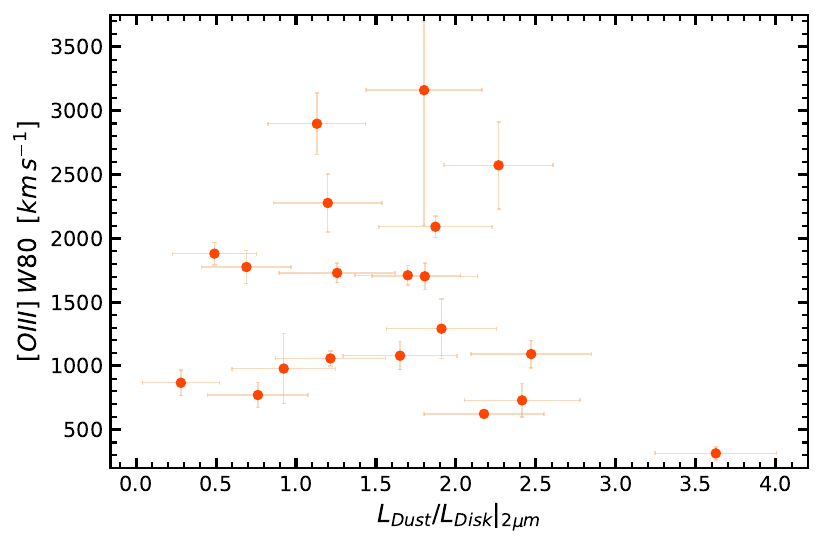} 
 \caption{The hot dust amplitude versus [\ion{O}{iii}] 80 per cent velocity widths \citep{2019MNRAS.487.2594T} for HRQs. Using the [\ion{O}{iii}] velocity widths as a probe for narrow-line region winds we see no correlation between torus-scale dust and kiloparsec scale outflow velocities in HRQs.}
 \label{fig:Ledd_w80}
\end{figure}

Unlike HRQs, quasars suffering only modest dust attenuation - e.g. red quasi-stellar objects \citep[rQSOs;][$\rm E(B-V) \simeq 0.12$]{2019MNRAS.488.3109K} - have remarkably similar multi-wavelength SEDs to their blue counterparts \citep{2021A&A...649A.102C}. Fig. \ref{fig:Ledd_w80} shows the [\ion{O}{iii}]\,$\lambda5008$\AA\ 80 per cent velocity widths \citep[calculated in;][]{2019MNRAS.487.2594T,2024MNRAS.533.2948S} versus the hot dust amplitude for HRQs. The [\ion{O}{iii}] velocity width can be used as a probe for narrow-line region (NLR) winds \citep{2005MNRAS.358.1043B} and depends strongly on the luminosity and therefore Eddington ratio of quasars with little dependence on the line-of-sight extinction \citep{2019MNRAS.487.2594T,2020A&A...634A.116V}. Consistent with these results we find no dependence of the [\ion{O}{iii}] velocity widths on $\rm E(B-V)$ and also no dependence on the hot dust amplitude as shown in Fig. \ref{fig:Ledd_w80} - contrary to the positive correlation between hot dust emission and outflow velocity observed in blue SDSS quasars and rQSOs \citep{2021MNRAS.501.3061T,2021A&A...649A.102C}. This suggests that the HRQ population is distinct from rQSOs, with rQSOs most likely tracing the red tail of the blue SDSS quasar population rather than the "blow-out" phase associated with merger-driven galaxy growth.

While HRQs host significant NLR winds, the wind velocities do not appear to be directly connected to the depletion of the hottest dust in the torus. Assuming that HRQs are tracing quasars that have blown away their torus-scale dust, this may suggest that the timescales upon which the torus-scale dust is depleted via feedback are significantly different to that of the NLR winds. In less obscured sources, parsec- and kiloparsec-scale wind velocities both correlate with each other as well as with the hot dust emission \citep{2019MNRAS.486.5335C,2021MNRAS.501.3061T}. Should torus-scale dust be ejected in a short-lived post-merger "blow-out" phase, the reduced opacity could result in the stalling of broad-line region (BLR) winds. However, if the NLR winds operate on longer timescales than the hot dust depletion, the connection between the hot-dust amplitude and NLR wind velocity could be broken, explaining the results presented in Fig. \ref{fig:Ledd_w80}.

\subsubsection{Hot-dust poor AGN at cosmic noon and beyond}

Hot-dust poor quasars have previously been linked to an early stage in quasar evolution where the dusty torus is yet to develop. However, searches for dust-poor quasars have focused on unobscured populations. For example, \citet{2010Natur.464..380J,2013ApJ...779..104J} invoked the following selection; 

\begin{equation}
    \begin{aligned}
        F_{2.3} = log_{10}\{\lambda F_\lambda|_{2.3\,\mu m}/ \lambda F_\lambda|_{0.51\,\mu m}\}<-0.5
        \label{eq:dust_selection_blue}
    \end{aligned}
\end{equation}

\noindent By de-reddening the SEDs of the 50 HRQs whose redshifts $z_{\rm sys}\geq1.5$ and $\overline{\chi^2_\nu} < 5.0$, we can test how many HRQs would be considered hot-dust poor by the \citet{2013ApJ...779..104J} formalism, and for which values of $\rm L_{Dust}/L_{Disk}|_{2\mu m}$ the \citet{2013ApJ...779..104J} condition is satisfied. We find that 7/50 or equivalently 14 per cent of the HRQ sample meet the selection described by Eqn. \ref{eq:dust_selection_blue} - corresponding to hot dust amplitudes $\rm L_{Dust}/L_{Disk}|_{2\mu m}\lesssim 0.75$. The mean of the distribution in $F_{2.3}$ is $\langle F_{2.3}\rangle|_{\rm HRQ} = -0.24\pm0.19$ for the HRQ population, placing the \citet{2013ApJ...779..104J} condition for dust poorness just 1$\sigma$ from the sample mean.

\subsection{Connecting dust emission to extinction and scattering}

\begin{figure*}
\centering
\begin{tabular}{cc}
 \includegraphics[scale=0.54]{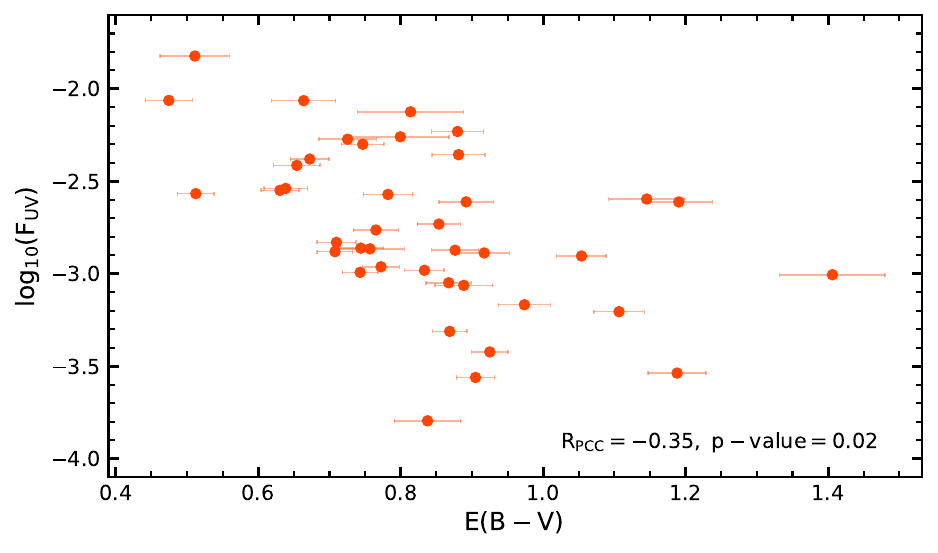} & \includegraphics[scale=0.54]{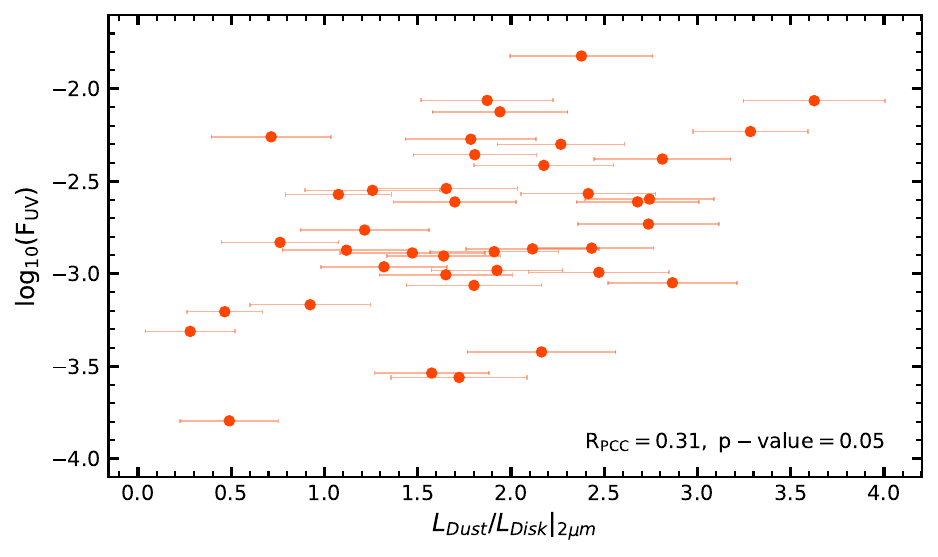} \\
 \end{tabular}
  \caption{The rest-frame ultraviolet scattering fraction, $F_{\rm UV}$ as a function of dust extinction, $\rm E(B-V)$ (left) and hot dust amplitudes, $\rm L_{Dust}/L_{Disk}|_{2\mu m}$ (right). We see a negative correlation with $\rm E(B-V)$ and a positive correlation with the hot dust amplitude.}
 \label{fig:FUV_EBV_dust}
\end{figure*}

Having conducted a multi-wavelength SED fit for a large HRQ population, which spans from the rest-frame UV to the rest-frame infrared, we can look for links between the dust attenuation that affects the bluer wavelengths and the dust emission in the infrared. Should the primary source of the UV excess in HRQs be scattered emission from the obscured quasar, we would expect the scattering fraction to depend on the properties of the dust which acts as the scattering medium. \autoref{fig:FUV_EBV_dust} illustrates that the scattering fraction, $F_{\rm UV}$, is weakly anti-correlated with the extinction suffered by the quasar continuum, $\rm E(B-V)$ with a Pearson Correlation Coefficient $R_{PCC} = -0.35$ and a $p$-value = 0.02. Hence, the UV continuum luminosities of HRQs are generally fainter in the more heavily extincted sources. As discussed in Section \ref{sec:dust}, we can use the sublimation-temperature dust amplitude of a quasar to measure the amount of dust present on 10s-of-pc scales \citep{2021MNRAS.501.3061T}. In Fig. \ref{fig:FUV_EBV_dust}, we see evidence of a weak positive correlation between the UV scattering fraction and the sublimation-temperature dust amplitudes with a Pearson Correlation Coefficient $R_{PCC} = 0.31$ and a $p$-value = 0.05.

The relations presented in Fig. \ref{fig:FUV_EBV_dust} suggest that the line-of-sight extinction is dominated by dust at larger scales, however, the extended dust is not primarily responsible for the scattering of quasar emission at rest-UV wavelengths. Instead, the correlation between the hot dust amplitude and the UV scattering fraction suggests that the scattering medium is primarily dominated by dust at the sublimation temperature. There are a number of dust geometries that enable hotter dust components to act as the primary scattering medium. For example, image polarimetry studies of blue Hot DOGs suggest that their outflow cones, also extending to kpc-scales, may be lined by a layer of inverse-sublimated graphite dust which can singly scatter sight-lines close to the boundary layer of the outflow towards the observer \citep{2022ApJ...934..101A,2025A&A...702A.124A}.  

Alternatively, spectropolarimetry studies of ERQs with near-Eddington accretion rates suggest that fast outflows (1000 $\rm km\;s^{-1}$) behave as a ‘skin’ near the edges of a geometrically thick disk and can scatter the intrinsic emission at 10s-of-pc scales \citep{2023MNRAS.525.2716Z}. The dispersion in Fig. \ref{fig:FUV_EBV_dust} can therefore be explained by these two competing geometries in addition to the varying host galaxy contribution to the rest-UV continuum across the HRQ sample - which can produce over-estimates of $F_{\rm UV}$. 

To more robustly test the trend observed in Fig. \ref{fig:FUV_EBV_dust}, we require polarimetry studies as well as a larger sample of dust-obscured quasars with a broader dynamic range in $F_{\rm UV}$ and a more complete distribution in sublimation-temperature dust amplitudes. By extrapolating the relationship in Fig. \ref{fig:i-band_selections}, we estimate that by amending the $(i-K)_{\rm AB}$ colour selection criteria to $(i-K)_{\rm AB}>0$ mag, HRQs with scattering fractions $F_{\rm UV}\sim 2$ per cent would no longer escape the selection and facilitate a less biased analysis. Furthermore, by relaxing the mid-infrared colour selection to $(W1-W2) >0.5$ mag, corresponding to the 90 per cent completeness limit of the \textit{WISE} AGN sample from \citet{Assef18:WISE}, we estimate we will be more complete to the full range of HRQs including those that are hot-dust poor at $z_{\rm sys}\geq2$, though at the expense of higher contamination from non-AGN populations to the colour selection. 

\subsection{Comparison to Little Red Dots}

In Section \ref{sec:UVExcess}, we conclude that excess emission at rest-UV wavelengths leading to an upwards inflection in the blue SED is extremely common in the HRQ sample. We tested two potential sources of the UV excess; - (i) a multi-component SED model where the UV excess is described as scattered quasar emission (and/or star formation in the quasar host) and (ii) a single-component SED model where the UV excess is described as an intrinsic feature of the rest-UV/optical continuum as a result of e.g. atypical dust properties. A key result from Section \ref{sec:UVExcess} is that none of the HRQs favoured the single-component SED model.

Conversely, the \emph{JWST} LRD population, which also host red rest-optical continua in addition to blue rest-UV photometric colours, seem to favour descriptions of the UV excess that describe the "V-shaped" rest-UV/optical continuum as an intrinsic feature. For example, the Orion dust-extinction law, discussed in Section \ref{sec:orion}, can reproduce the SEDs of LRDs without the need for an additional scattered/stellar emission component \citep{2025ApJ...980...36L}. In addition, strong Balmer absorption is seen in some LRDs, explained by an AGN accretion disk embedded in dense neutral gas clumps, effectively absorbing the continuum emission blueward of the Balmer break \citep[e.g.][]{2025ApJ...980L..27I,2025arXiv250316600D,2025arXiv250316596N}. 

\begin{table*}
    \centering
    \caption{The \textsc{qsogen} best-fit parameters for the 4 HRQs which meet the \citet{2024arXiv240403576K} LRD selection criteria as well as their black hole masses from previous works - i.e. \citet{2012MNRAS.427.2275B,2013MNRAS.429L..55B,2015MNRAS.447.3368B,2019MNRAS.487.2594T,2024MNRAS.533.2948S} - and their Eddington-scaled accretion rates. Extinction-corrected luminosities are given in $\rm erg\;s^{-1}$, full-width at half-maximum's (FWHMs) are given in $\rm  km\,s^{-1}$, and black hole masses are given in solar masses ($\rm M_\odot$).}    
    \begin{tabular}{|l|c|c|c|c|c|c|c|c|}
         \hline
          Object & $z_{\rm sys}$ &log$_{10}\{\lambda$L$_{\lambda} (3000\mathring{A})\}$&$\rm E(B-V)$&F$_{\rm{UV}}$ [$\%$] &$\rm L_{Dust}/L_{Disk}|_{2\mu m}$& FWHM (H$\alpha$) &log$_{10}$(M\textsc{bh}) & log$_{10}(\lambda_{\rm Edd})$ \\
         \hline
            ULASJ2315+0143 & 2.566 & 47.69$\pm$0.08 & 1.40$\pm$0.08 & 0.10$\pm$0.02 & 1.68$\pm$0.35 & 4040 & 10.17 & -0.41 \\
            VIKJ2230-2956  & 1.319 & 46.08$\pm$0.03 & 1.63$\pm$0.13 & 0.36$\pm$0.03 & 6.06$\pm$0.36 & 2409 & 8.63 & -0.07 \\
            VIKJ2243-3504  & 2.085 & 47.32$\pm$0.05 & 1.19$\pm$0.05 & 0.24$\pm$0.03 & 1.72$\pm$0.31 & 3197 & 8.99 & 0.57 \\
            VIKJ2256-3114  & 2.329 & 46.85$\pm$0.04 & 0.88$\pm$0.04 & 0.45$\pm$0.03 & 1.91$\pm$0.36 & 4074 & 9.16 & 0.02 \\
         \hline
    \end{tabular}
    \label{tab:hrq_lrd}
\end{table*}

Given the qualitative similarities between the UV/optical SEDs observed in HRQs and LRDs, HRQs could trace hyper-luminous analogues of the LRD population at cosmic noon \citep{2024MNRAS.533.2948S}. However, given their different selections, further exploration of the similarities and differences between the SED properties of HRQs and LRDs is now required to understand whether the two populations are tracing similar physics in their respective red AGN phases. There have been multiple selection criteria employed to identify LRDs at both very high redshifts \citep[e.g. $z_{\rm sys}>5$][]{2024arXiv240403576K} and extending now to lower redshifts approaching cosmic noon \citep{Ma:25,Euclid:25} where their number densities appear to decline. The LRD selection criteria described in \citet{2024arXiv240403576K} employs the continuum blue-ward and red-ward of the 3645\AA\, break to define the LRD population. To determine which of the 51 HRQs for which we have rest-UV coverage meet the \citet{2024arXiv240403576K} criteria, their continua are split at 3645\AA\, in the rest frame and fit with a least-squares minimisation of Eqn. \ref{eq:LRD_Selection_model};

\begin{equation}
    \begin{aligned}
      m_i = -2.5\, \beta\, {log}_{10}(\lambda_i) + c\\
    \end{aligned}  
    \label{eq:LRD_Selection_model} 
\end{equation}

\noindent where $m_i$ is the AB magnitude of the $i^{\rm th}$ filter, $\beta$ is the continuum slope, $\lambda_i$ is the effective wavelength of the $i^{\rm th}$ filter in $\mu m$ and $c$ is a normalisation constant. An object is formally considered an LRD if their continuum slopes meet the following criteria \citep{2024arXiv240403576K};

\begin{enumerate}
    \item $\beta_{\rm opt}>0$
    \item $-2.8< \beta_{\rm uv} < -0.37$
\end{enumerate}

By applying these criteria to the HRQ sample we find that 4/51 $z>1.5$ HRQs formally satisfy the constraints. Despite most sources (47/51) hosting sufficiently red optical continua, HRQs are insufficiently blue at rest-UV wavelengths to be considered an LRD in this formalism - consistent with the composite SED shapes of HRQs and LRDs shown in Fig. \ref{fig:Stacked_Phot_HRQ}. The average $(i-K)$ colour of the LRD composite is 2.8 mag, which is very close to the HRQ selection threshold of $>$2.5. LRD-like HRQs with blue UV continuum slopes are therefore removed from the HRQ sample by our $(i-K)_{\rm AB}$ colour selection meaning that the difference in the SED shape between the two populations is largely a selection effect. Nevertheless, it is interesting that despite the pre-\textit{JWST} HRQ selection criteria biasing against sources with ``V-shaped" SEDs, we find 4 of the HRQs meet all the main LRD selection criteria and are therefore bonafide cosmic noon analogues of LRDs. The properties of these 4 sources are summarised in Table \ref{tab:hrq_lrd} including The Big Red Dot - J2315 - studied in detail in \citet{2024MNRAS.533.2948S}. 

\begin{figure}
\centering
 \includegraphics[scale=0.65, trim={0.0cm, 0.25cm, 0.0cm, 0.2cm}, clip]{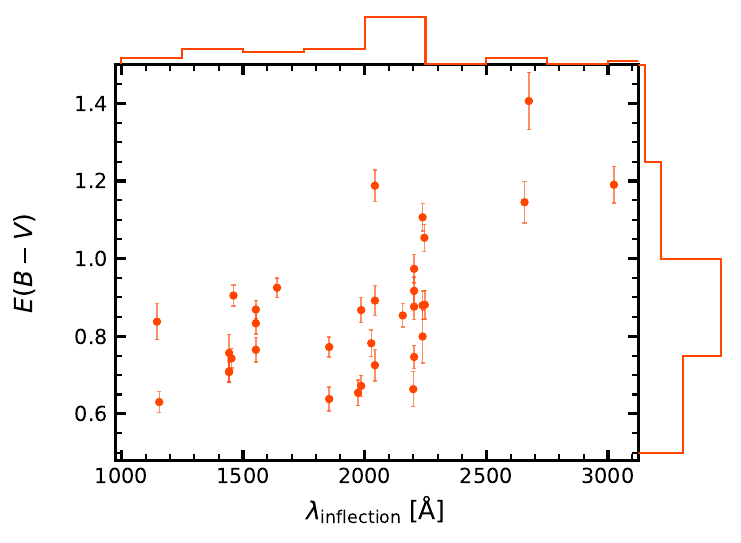} 
  \caption{The UV/optical inflection point wavelength as a function of $\rm E(B-V)$ for the HRQ sample. The HRQ inflection points are inconsistent with the Balmer Break, $\lambda_{\rm rest}= 3645$\AA,\, and have a broad distribution (histogram, top).}
 \label{fig:Inflection_EBV}
\end{figure}

To better understand why most of the HRQs do not satisfy the LRD selection in \citet{2024arXiv240403576K}, in Fig. \ref{fig:Inflection_EBV}, we present the distribution of the rest-UV/optical inflection points for the 42 HRQ SEDs in which a statistically significant UV excess was observed. We find an average UV/optical inflection point $\langle\lambda_{\rm rest}\rangle = 2156 \pm 493$\AA\, in the HRQ sample, which is significantly bluer than the Balmer Break. Thus, it is extremely unlikely that Balmer absorption is the primary source of the UV excess in HRQs. 

The large range in UV/optical inflection points in the HRQ sample is what drives the success of the multi-component SED model proposed in Section \ref{sec:SED_Fits}. Multi-component SED models have the flexibility to reproduce a variety of UV/optical inflection points because, in this framework, the inflection point of a given SED depends on the relative contributions of the scattered and dust attenuated components. Conversely, in single-component models where some intrinsic quality, like Balmer absorption, is the primary cause of the UV excess, the inflection point is determined by the Balmer limit, inconsistent with the dust-extinction dependent inflection points observed in HRQs. It is therefore likely that HRQs and LRDs despite having similar SED shapes in the rest-UV/optical, are fundamentally different in terms of the reason for their excess rest-UV emission. For example, the dense neutral gas, in which the accretion disks of LRDs are thought to be embedded, can host very low metallicities \citep[e.g.][]{2025arXiv250316600D,2025arXiv250316596N}, while HRQs are iron rich \citep[e.g.][]{2019MNRAS.487.2594T,2024MNRAS.533.2948S}. However, more recent results have confirmed an abundance of \ion{Fe}{ii} emission in several LRDs \citep[e.g.][]{Labbe:25,2025arXiv251000103T,2025arXiv251107515K}, emphasising the diversity of the LRD population and high-lighting the parallels between LRDs and luminous red quasars.

While LRDs and HRQs share some similarities in terms of their SEDs - namely excess UV emission, a lack of hot dust and evidence for super-Eddington accretion rates, they also probe very different regions in the AGN luminosity and black hole mass functions with the HRQs hosting black holes that are several orders of magnitude larger than in LRDs. The population densities of LRDs also appear to decline at $z<4$ \citep{2024arXiv240403576K} and attempts to find LRDs over wider areas than covered by \textit{JWST} have uncovered relatively few luminous examples \citep{Ma:25b}. HRQs are significantly rarer than LRDs on account of their very high luminosities and black hole masses, however, they potentially dominate unobscured, blue quasars at similar luminosities and redshifts \citep{2015MNRAS.447.3368B}. Future wide-field surveys for HRQs should broaden the selection criteria to include candidates with bluer UV slopes and less hot dust to ascertain whether LRD-like SEDs are indeed common in the most luminous, reddened quasars at cosmic noon. 

\section{Conclusions} \label{SEC:HRQ_CONC}

We have conducted spectral energy distribution (SED) fitting to a complete sample of 63 spectroscopically confirmed heavily reddened quasars (HRQs) at $0.7\lesssim z_{\rm sys}\lesssim2.7$ to model their rest-frame UV to infrared emission. We utilise multi-wavelength surveys spanning the $ugriz-YJHK-W1W2$ filters, which corresponds to a wavelength coverage of 1000\AA-3$\mu m$ at $z_{\rm sys}\sim2$. We show that in most cases a two-component SED model featuring a dust-attenuated quasar component and an additional blue scattered light component, provides a good fit to the photometric data. Our main findings are as follows;

\begin{itemize}

    \item We find that HRQs have significantly higher luminosities and accretion rates compared to blue quasars, with Eddington ratios approximately an order of magnitude larger than that of their blue SDSS counterparts at similar redshift.
    
    \item 51 of the 63 HRQs have sufficient photometric data blueward of the Balmer break to investigate their UV shapes. We confirm that blue photometric colours at rest-UV wavelengths are ubiquitous in the HRQ sample, with a UV excess consistent with scattered quasar emission detected in 42/51 of the HRQs. Conversely just 4/51 HRQs do not feature a statistically significant UV excess and the remaining 5/51 yielded inconclusive results. We calculate an average scattered fraction - $\langle F_{\rm{UV}}\rangle_{\rm HRQs} = 0.26$ per cent. While comparatively lower than other red AGN populations, this is most likely due to selection effects with the $(i-K)_{\rm AB}>2.5$ colour selection for HRQs removing quasars with $F_{\rm{UV}}\gtrsim1$ per cent from the sample.

    \item We find that a single-component SED model using the Orion Nebula dust-extinction law, successfully used by \citet{2025ApJ...980...36L} to fit the SEDs of LRDs, is not favoured over the multi-component SED model used in this work for any of the HRQs. This is because the Orion dust-extinction law is too flat to reproduce the rest-optical continuum emission of HRQs. 

    \item When restricting the HRQ sample to redshifts, $z_{\rm sys}\geq1.5$, we find that the distribution of hot dust amplitudes in the HRQ sample is biased towards lower values than that of blue SDSS quasars, with $\langle \rm L_{Dust}/L_{Disk}|_{2\mu m} \rangle = 1.6\pm0.8$ and $4.4\pm2.1$ for HRQs and blue SDSS quasars, respectively. Given that our mid-infrared, \textit{WISE} selection criteria may bias the HRQ sample to higher dust amplitudes at more modest dust extinctions, there may exist more HRQs with even lower hot dust amplitudes than observed in this sample. We interpret the deficit in hot emission as evidence that HRQs are observed in a radiative feedback dominated "blowout" phase, where equatorial dust on torus scales has been depleted. Alternatively, the line-of-sight extinctions from the SED fitting may be under-estimated with the hottest components of the torus 'self-absorbed' by more significant dust columns. We favour the former scenario given the already extreme dust-corrected luminosities and the Eddington/super-Eddington accretion rates observed in HRQs. 

    \item We find that the UV scattering fractions of HRQ are weakly correlated with the hot dust amplitude and anti-correlated with the line-of-sight extinction. This suggests that dust at the sublimation temperature is likely acting as the scattering medium, potentially distributed along the boundary of extended bipolar outflow cones. However, given the high line-of-sight extinction and lack of torus-scale dust in HRQs, the obscuring medium is most likely dominated by dust located on larger scales in the host galaxy interstellar medium. 

    \item We find that while HRQs host significant [\ion{O}{iii}] winds \citep{2019MNRAS.487.2594T}, their outflow velocities are not correlated with their hot dust emission properties. We interpret this as evidence that the timescales on which torus-scale dust is depleted via feedback are different to that of the narrow-line region winds.
    
    \item We look at similarities and differences between the average SEDs of HRQs and the \textit{JWST} LRD population. While 47/51 HRQs with blue enough wavelength coverage have a sufficiently red optical continuum to satisfy the LRD selection in \citet{2024arXiv240403576K}, only 4/51 are sufficiently blue in the rest-UV. We largely attribute this to a selection effect as HRQs were selected to have red $(i-K)$ colours and pre-date the discovery of LRDs. Nevertheless, the HRQ population contains four bonafide, cosmic noon massive analogues of LRDs. We find that descriptions of the UV excess that invoke some intrinsic property of the SED as the primary cause, such as Balmer absorption, are inconsistent with the SEDs of HRQs. Instead, the rest-UV/optical inflection points of HRQ SEDs are diverse, with a mean inflection point wavelength, $\langle\lambda_{\rm rest}\rangle = 2156 \pm 493$\AA.\, We interpret the diversity in the SED inflection point wavelength as evidence that a multi-component SED model is the best description of the UV excess in HRQs and Balmer absorption is unlikely the primary origin of the "V-shaped" continuum. We also show that, similar to LRDs, HRQs are hot-dust poor, however, at least some HRQs have significant cold dust reservoirs, unlike LRDs.

\end{itemize}

Overall our comprehensive investigation of the full UV to infrared SEDs of a complete sample of HRQs has unearthed important similarities and differences with other obscured AGN populations such as Hot DOGs and LRDs as well as enabling the physical properties and geometry of the dust in these quasars to be constrained. We find that massive analogues of the LRD population do exist at cosmic noon. Future imaging and spectroscopic surveys should widen the HRQ selection criterion to include sources with bluer continuum slopes and flatter hot dust SEDs in order to establish more complete samples from which their demographics can be constrained. 

\section*{Acknowledgements}

We thank Zhengrong Li for sharing the Orion dust extinction curve used in this analysis, Prof. Roberto Assef for their invaluable insights into Hot DOGs and the anonymous referee for reviewing this work. MS acknowledges funding from the University of Southampton via the Mayflower studentship. MB and ST acknowledge funding from the Royal Society via a University Research Fellowship Renewal Grant and Research Fellows Enhancement Award (URF/R/221103; RF/ERE/221053). MT acknowledges funding from the Science and Technology Funding Council (STFC) grant, ST/X001075/1.

\section*{Data Availability} 

The full catalogue of photometric data used for this analysis will be made available as online only supplementary material.

\bibliographystyle{mnras}
\bibliography{UOS} 

\appendix 

\section{The heavily reddened quasar "best-fit" SEDs} \label{App:plots}

The best-fit SED parameters for the 63 HRQs with their corresponding MCMC uncertainties. The 12 HRQs for which it was not possible to conduct a multi-component SED fit to model the UV excess are classified as "N/A". Objects for which a UV excess was confirmed, rejected or yielded inconclusive results are denoted as "Conf.", "Rej." or "Inc." respectively. HRQs with redshifts $z_{\rm sys}<1.5$ are modelled with $fragal=0.05$, otherwise $fragal=0.00$. The "best fit" models are also represented graphically in Fig. \ref{fig:SEDs1}.

\begin{table*}
    \centering
    \caption{The best-fit SED parameters and their corresponding MCMC uncertainties for 63 HRQs in addition to the Eddington ratios calculated in Section \ref{sec:properties}.}    
    \begin{tabular}{|l|c|c|c|c|c|c|c|c|}
        \hline
          Object & $z_{\rm sys}$ & log$_{10}(\lambda_{\rm Edd})$ & log$_{10}\{\lambda$L$_{\lambda} (3000\mathring{A}) [\rm erg\;s^{-1}]\}$&$\rm E(B-V)$ &F$_{\rm{UV}}$ [$\%$] &$\rm L_{Dust}/L_{Disk}|_{2\mu m}$&$\overline{\chi}^2_\nu$ & UV Excess \\
         \hline
            ULASJ0016-0038 & 2.194 & -0.19 & 46.63$\pm$0.05 & 0.51$\pm$0.03 & $-$ & 2.40$\pm$0.36 & 2.7 & Rej.            \\
            ULASJ0041-0021 & 2.517 & -0.02 & 47.27$\pm$0.04 & 0.87$\pm$0.02 & 0.05$\pm$0.01 & 0.35$\pm$0.22 & 3.3 & Conf. \\
            ULASJ0123+1525 & 2.629 &  0.50 & 48.12$\pm$0.09 & 1.74$\pm$0.09 & 0.05$\pm$0.01 & 2.06$\pm$0.37 & 5.9 & Conf. \\
            ULASJ0141+0101 & 2.562 & -0.44 & 46.61$\pm$0.05 & 0.48$\pm$0.03 & 0.86$\pm$0.04 & 1.94$\pm$0.39 & 4.4 & Conf. \\
            ULASJ0144-0114 & 2.505 & -0.03 & 47.31$\pm$0.10 & 0.84$\pm$0.04 & 0.11$\pm$0.01 & 1.96$\pm$0.29 & 1.3 & Conf. \\
            ULASJ0144+0036 & 2.283 & -0.24 & 47.14$\pm$0.10 & 0.84$\pm$0.04 & 0.02$\pm$0.01 & 0.52$\pm$0.29 & 1.6 & Conf. \\
            ULASJ0221-0019 & 2.247 &  0.36 & 46.91$\pm$0.04 & 0.74$\pm$0.02 & 0.10$\pm$0.02 & 2.48$\pm$0.27 & 1.0 & Conf. \\
            ULASJ1002+0137 & 1.595 &  0.32 & 46.79$\pm$0.06 & 1.11$\pm$0.07 & 0.06$\pm$0.01 & 0.43$\pm$0.26 & 2.1 & Conf. \\
            ULASJ1216-0313 & 2.574 & -0.16 & 48.76$\pm$0.05 & 1.14$\pm$0.04 & $-$ & 2.85$\pm$0.36 & 2.4 & N/A             \\
            ULASJ1234+0907 & 2.503 & -0.28 & 48.13$\pm$0.07 & 1.69$\pm$0.08 & $-$ & 1.15$\pm$0.34 & 8.0 & N/A             \\
            ULASJ1415+0836 & 1.120 & -0.57 & 45.95$\pm$0.16 & 1.43$\pm$0.09 & $-$ & 8.94$\pm$0.16 & 24.2 & N/A            \\
            ULASJ1455+1230 & 1.460 &  0.18 & 46.79$\pm$0.07 & 1.56$\pm$0.24 & $-$ & 1.65$\pm$0.28 & 2.6 & N/A             \\
            ULASJ1539+0557 & 2.658 &  0.25 & 48.03$\pm$0.07 & 1.26$\pm$0.06 & $-$ & 1.26$\pm$0.37 & 0.4 & N/A             \\
            ULASJ2200+0056 & 2.541 &  0.55 & 47.44$\pm$0.05 & 0.63$\pm$0.02 & 0.28$\pm$0.02 & 1.23$\pm$0.36 & 0.5 & Conf. \\
            ULASJ2224-0015 & 2.223 &  0.29 & 46.97$\pm$0.05 & 0.71$\pm$0.03 & 0.14$\pm$0.02 & 1.98$\pm$0.34 & 0.6 & Conf. \\
            ULASJ2312+0454 & 0.700 & -0.41 & 45.17$\pm$0.16 & 1.14$\pm$0.04 & 1.49$\pm$0.04 & 7.85$\pm$0.18 & 3.5 & Conf. \\
            ULASJ2315+0143 & 2.566 & -0.41 & 47.69$\pm$0.08 & 1.40$\pm$0.08 & 0.10$\pm$0.02 & 1.68$\pm$0.35 & 1.2 & Conf. \\
            VHSJ1117-1528  & 2.428 & -0.08 & 47.18$\pm$0.07 & 0.97$\pm$0.07 & $-$ & 1.25$\pm$0.35 & 1.1 & N/A             \\
            VHSJ1122-1919  & 2.464 &  0.35 & 47.36$\pm$0.06 & 0.87$\pm$0.06 & $-$ & 0.64$\pm$0.32 & 4.0 & N/A             \\
            VHSJ1301-1624  & 2.138 &  0.25 & 47.45$\pm$0.07 & 1.14$\pm$0.07 & $-$ & 1.07$\pm$0.27 & 1.2 & N/A             \\
            VHSJ1350-0503  & 2.176 &  $-$  & 47.18$\pm$0.05 & 0.77$\pm$0.03 & 0.26$\pm$0.01 & 1.04$\pm$0.35 & 0.8 & Conf. \\
            VHSJ1409-0830  & 2.300 &  $-$  & 47.13$\pm$0.07 & 0.94$\pm$0.08 & $-$ & 2.10$\pm$0.34 & 1.6 & N/A             \\
            VHSJ1556-0835  & 2.188 &  0.66 & 46.93$\pm$0.07 & 0.78$\pm$0.06 & 0.55$\pm$0.04 & 0.68$\pm$0.29 & 0.8 & Conf. \\
            VHSJ2024-5623  & 2.282 & -0.71 & 46.72$\pm$0.09 & 0.91$\pm$0.12 & $-$ & 0.76$\pm$0.30 & 0.4 & N/A             \\
            VHSJ2028-4631  & 2.464 & -0.19 & 46.59$\pm$0.05 & 0.49$\pm$0.03 & 1.51$\pm$0.06 & 2.42$\pm$0.29 & 0.6 & Conf. \\
            VHSJ2028-5740  & 2.121 & -0.28 & 47.66$\pm$0.06 & 1.21$\pm$0.04 & 0.03$\pm$0.01 & 1.57$\pm$0.33 & 3.1 & Conf. \\
            VHSJ2048-4644  & 2.182 & -0.06 & 46.79$\pm$0.05 & 0.89$\pm$0.01 & 0.86$\pm$0.06 & 2.87$\pm$0.29 & 1.3 & Conf. \\
            VHSJ2100-5820  & 2.360 &  0.26 & 47.08$\pm$0.06 & 0.90$\pm$0.03 & 0.03$\pm$0.01 & 1.58$\pm$0.31 & 1.9 & Conf. \\
            VHSJ2101-5943  & 2.313 & -0.97 & 47.27$\pm$0.09 & 0.63$\pm$0.03 & $-$ & 2.41$\pm$0.33 & 1.9 & Rej.            \\
            VHSJ2109-0026  & 2.344 & -0.77 & 46.66$\pm$0.04 & 0.64$\pm$0.03 & 0.29$\pm$0.02 & 1.76$\pm$0.34 & 2.9 & Conf. \\
            VHSJ2115-5913  & 2.115 &  0.39 & 47.49$\pm$0.05 & 1.07$\pm$0.04 & 0.12$\pm$0.01 & 1.60$\pm$0.32 & 1.5 & Conf. \\
            VHSJ2130-4930  & 2.448 & -0.03 & 47.21$\pm$0.06 & 0.92$\pm$0.04 & 0.13$\pm$0.02 & 1.47$\pm$0.37 & 2.2 & Conf. \\
            VHSJ2141-4816  & 2.655 &  0.00 & 47.38$\pm$0.05 & 0.87$\pm$0.03 & 0.14$\pm$0.01 & 1.14$\pm$0.38 & 0.6 & Conf. \\
            VHSJ2143-0643  & 2.383 & -0.75 & 46.93$\pm$0.08 & 0.81$\pm$0.09 & 0.75$\pm$0.04 & 1.99$\pm$0.28 & 0.4 & Inc.  \\
            VHSJ2144-0523  & 2.152 & -0.76 & 46.81$\pm$0.07 & 0.82$\pm$0.08 & $-$ & 2.26$\pm$0.35 & 1.7 & N/A             \\
            VHSJ2212-4624  & 2.141 & -0.70 & 46.87$\pm$0.06 & 0.97$\pm$0.07 & $-$ & 2.52$\pm$0.37 & 3.4 & N/A             \\
            VHSJ2220-5618  & 2.220 &  0.00 & 47.74$\pm$0.04 & 0.92$\pm$0.02 & 0.04$\pm$0.01 & 2.19$\pm$0.37 & 1.3 & Conf. \\
            VHSJ2227-5203  & 2.656 & -0.58 & 47.15$\pm$0.08 & 0.76$\pm$0.04 & 0.13$\pm$0.02 & 2.14$\pm$0.36 & 7.9 & Conf. \\
            VHSJ2235-5750  & 2.246 & -0.82 & 46.99$\pm$0.04 & 0.67$\pm$0.03 & 0.41$\pm$0.03 & 2.44$\pm$0.34 & 2.7 & Conf. \\
            VHSJ2256-4800  & 2.250 & -0.60 & 47.23$\pm$0.05 & 0.85$\pm$0.03 & 0.19$\pm$0.02 & 2.63$\pm$0.35 & 3.1 & Conf. \\
            VHSJ2257-4700  & 2.156 & -0.44 & 46.69$\pm$0.05 & 0.74$\pm$0.03 & 0.13$\pm$0.03 & 2.42$\pm$0.36 & 0.3 & Inc.  \\
            VHSJ2306-5447  & 2.372 & -0.76 & 46.93$\pm$0.05 & 0.78$\pm$0.03 & 0.11$\pm$0.01 & 1.37$\pm$0.34 & 1.1 & Conf. \\
            VHSJ2332-5240  & 2.450 & -0.50 & 46.64$\pm$0.04 & 0.73$\pm$0.04 & 0.53$\pm$0.01 & 1.79$\pm$0.35 & 3.7 & Conf. \\
            VHSJ2355-0011  & 2.531 & -0.58 & 47.29$\pm$0.05 & 0.90$\pm$0.04 & 0.24$\pm$0.02 & 2.68$\pm$0.34 & 0.8 & Conf. \\
            VIKJ2205-3132  & 2.307 &  0.17 & 46.57$\pm$0.05 & 0.71$\pm$0.03 & 0.15$\pm$0.02 & 0.70$\pm$0.30 & 4.0 & Conf. \\
            VIKJ2214-3100  & 1.069 & -0.52 & 45.72$\pm$0.04 & 1.47$\pm$0.12 & 0.25$\pm$0.03 & 7.37$\pm$0.34 & 14.5 & Inc. \\
            VIKJ2228-3205  & 2.364 &  0.11 & 46.13$\pm$0.06 & 0.66$\pm$0.04 & 0.39$\pm$0.03 & 2.14$\pm$0.37 & 0.6 & Conf. \\
            VIKJ2230-2956  & 1.319 & -0.07 & 46.08$\pm$0.03 & 1.63$\pm$0.13 & 0.36$\pm$0.03 & 6.06$\pm$0.36 & 6.6 & Conf. \\
            VIKJ2232-2844  & 2.292 &  0.16 & 46.92$\pm$0.03 & 0.89$\pm$0.06 & $-$ & 1.76$\pm$0.26 & 0.9 & Rej.            \\
            VIKJ2238-2836  & 1.231 & -0.35 & 46.18$\pm$0.03 & 1.45$\pm$0.08 & 0.50$\pm$0.03 & 5.09$\pm$0.32 & 3.0 & Inc.  \\
            VIKJ2241-3006  & 0.720 &  0.40 & 45.33$\pm$0.06 & 1.52$\pm$0.23 & 0.38$\pm$0.04 & 7.30$\pm$0.33 & 4.8 & Conf. \\
            VIKJ2243-3504  & 2.085 &  0.57 & 47.32$\pm$0.05 & 1.19$\pm$0.05 & 0.25$\pm$0.03 & 1.65$\pm$0.31 & 1.1 & Conf. \\
            VIKJ2245-3516  & 1.335 & -1.09 & 46.02$\pm$0.03 & 1.40$\pm$0.09 & $-$ & 12.72$\pm$0.41 & 9.3 & Rej.           \\
            VIKJ2251-3433  & 1.693 & -0.36 & 46.48$\pm$0.05 & 1.15$\pm$0.17 & 0.27$\pm$0.03 & 2.74$\pm$0.36 & 4.6 & Conf. \\
            VIKJ2256-3114  & 2.329 &  0.02 & 46.85$\pm$0.04 & 0.88$\pm$0.04 & 0.45$\pm$0.03 & 1.91$\pm$0.36 & 0.7 & Conf. \\
            VIKJ2258-3219  & 0.879 &  0.80 & 45.78$\pm$0.04 & 1.46$\pm$0.15 & 0.39$\pm$0.04 & 5.82$\pm$0.32 & 3.7 & Conf. \\
            VIKJ2306-3050  & 1.060 & -0.24 & 45.51$\pm$0.05 & 1.44$\pm$0.16 & 0.25$\pm$0.04 & 3.40$\pm$0.32 & 3.6 & Inc.  \\
            VIKJ2309-3433  & 2.159 & -0.79 & 46.12$\pm$0.04 & 0.68$\pm$0.04 & 0.88$\pm$0.04 & 3.69$\pm$0.35 & 1.6 & Conf. \\
            VIKJ2313-2904  & 1.851 & -0.73 & 46.38$\pm$0.04 & 0.87$\pm$0.08 & 0.58$\pm$0.04 & 3.35$\pm$0.35 & 2.8 & Conf. \\
            VIKJ2314-3459  & 2.325 &  0.61 & 46.66$\pm$0.04 & 0.75$\pm$0.03 & 0.50$\pm$0.03 & 2.32$\pm$0.34 & 1.7 & Conf. \\
            VIKJ2323-3222  & 2.191 &  0.21 & 47.02$\pm$0.06 & 0.97$\pm$0.04 & 0.07$\pm$0.01 & 0.86$\pm$0.31 & 2.3 & Conf. \\
            VIKJ2350-3019  & 2.324 &  0.42 & 46.61$\pm$0.05 & 0.76$\pm$0.03 & 0.17$\pm$0.02 & 1.25$\pm$0.36 & 1.3 & Conf. \\
            VIKJ2357-3024  & 1.129 & -0.34 & 45.77$\pm$0.05 & 1.38$\pm$0.15 & 0.24$\pm$0.03 & 4.47$\pm$0.37 & 4.5 & Conf. \\  
         \hline
    \end{tabular}
    \label{Tab:SED_Results}
\end{table*}

\begin{figure*}
\centering
 \includegraphics[scale=0.82, trim={0.0cm, 2.5cm, 0.0cm, 2.0cm}, clip]{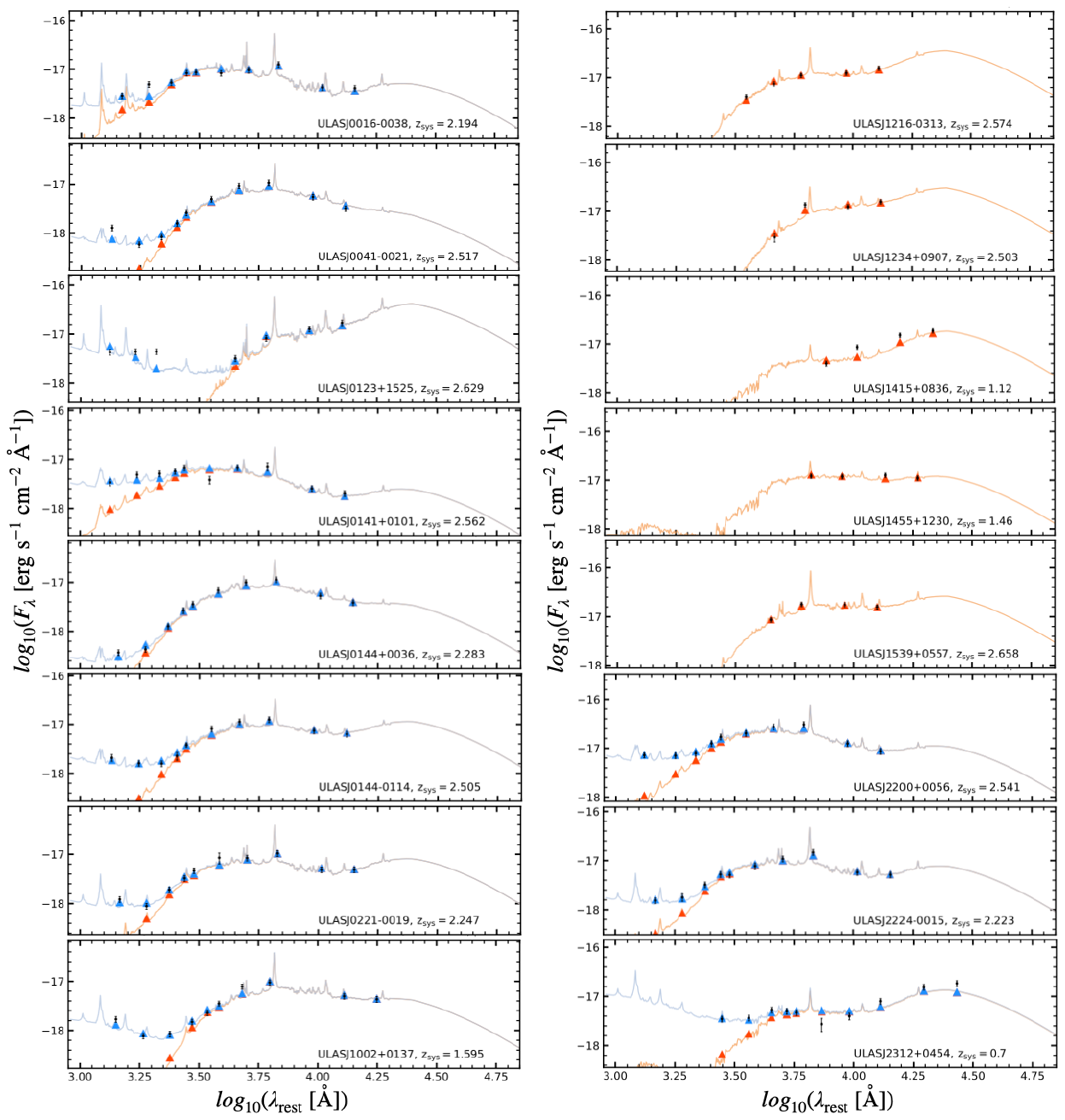} 
  \caption{The "best-fit" HRQ SEDs for all 63 sources. The photometric data and associated uncertainties are indicated in black. Where rest-UV photometry is available, the best-fit multi-component SED models are shown by the blue lines and triangles - the dust-attenuated quasar component is also presented in orange. Where rest-UV photometry is unavailable, the best-fit single-component SED models are represented by orange lines and triangles.}
 \label{fig:SEDs1}
\end{figure*}

\begin{figure*}
\ContinuedFloat
\centering
 \includegraphics[scale=0.82, trim={0.0cm, 2.5cm, 0.0cm, 2.0cm}, clip]{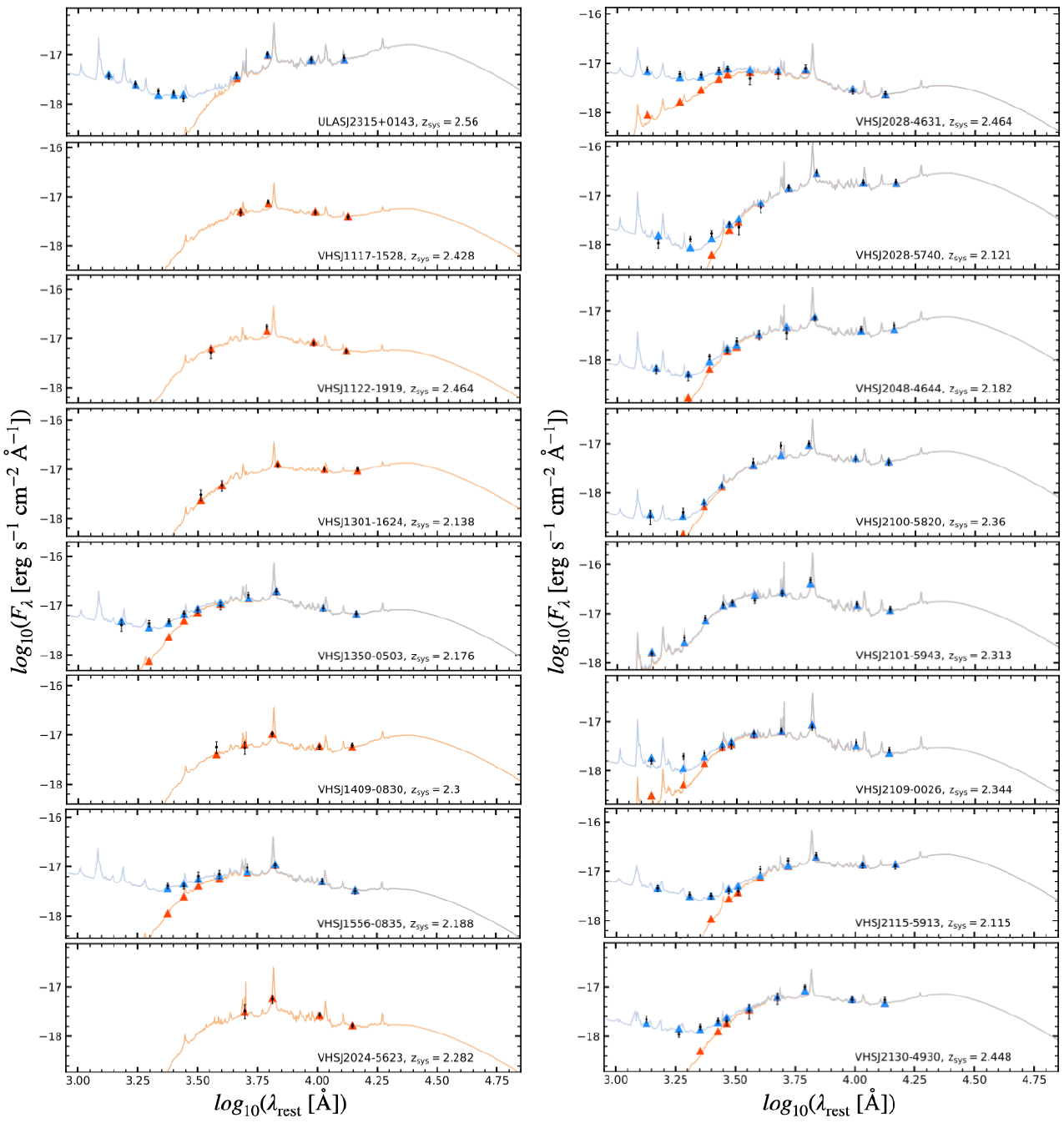} 
  \caption{\emph{(continued).}}
\end{figure*}

\begin{figure*}
\ContinuedFloat
\centering
 \includegraphics[scale=0.82, trim={0.0cm, 2.5cm, 0.0cm, 2.0cm}, clip]{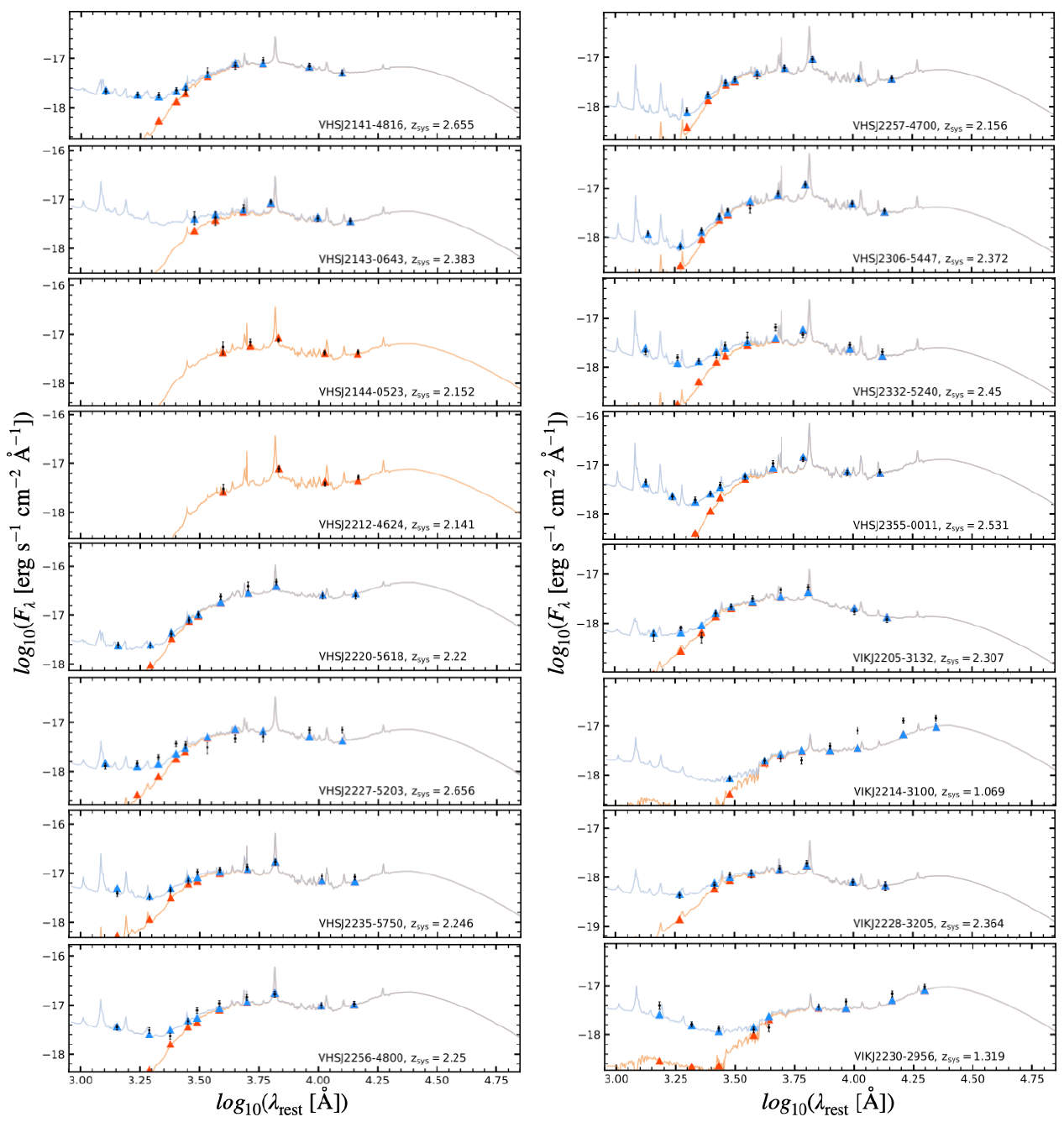} 
  \caption{\emph{(continued).}}
\end{figure*}

\begin{figure*}
\ContinuedFloat
\centering
 \includegraphics[scale=0.82, trim={0.0cm, 2.5cm, 0.0cm, 2.0cm}, clip]{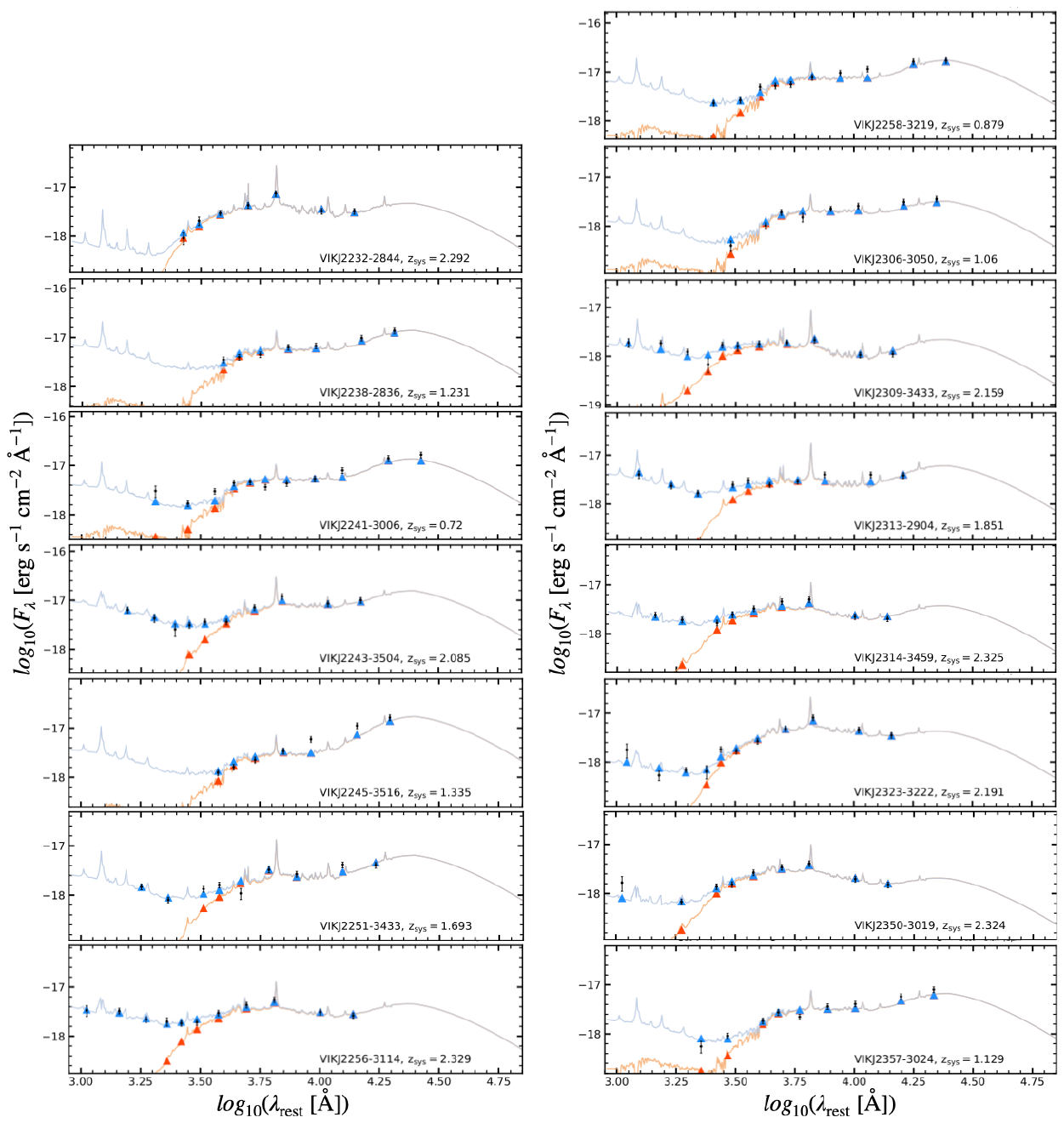} 
  \caption{\emph{(continued).}}
\end{figure*}

\clearpage 

\section{Orion-like dust properties as the source of the UV excess} \label{App:orion}

\begin{figure}
\centering
 \includegraphics[scale=0.3, trim={4.0cm, 1.0cm, 4.0cm, 0.0cm}]{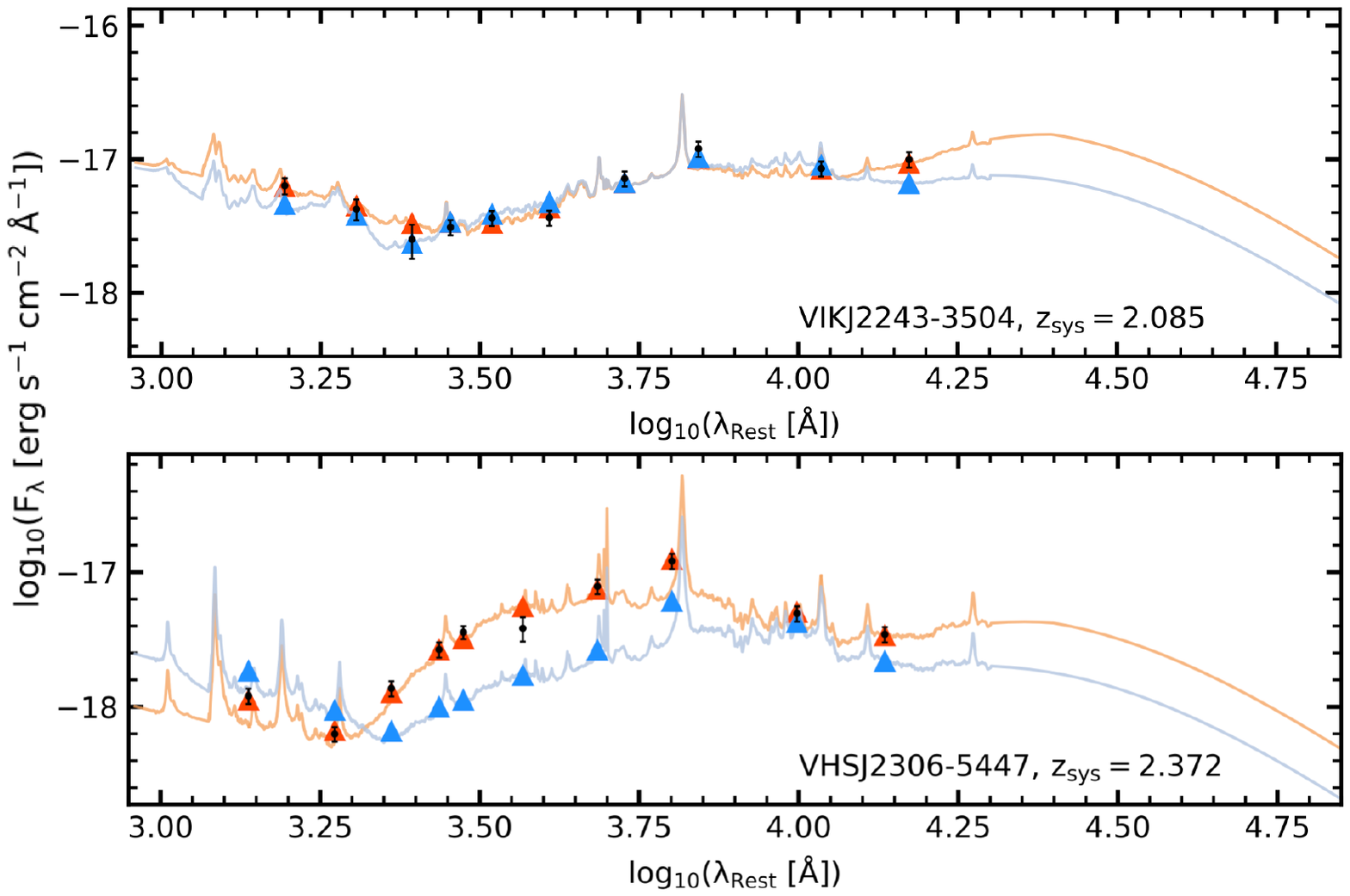} 
  \caption{SED fits for the HRQs; VIKJ2115-5913 (top) and ULASJ0144-0114 (bottom). The photometric data and associated uncertainties are presented in black. The best-fit SED models using the Orion dust law are presented by the blue lines and triangles. The best-fit SED models using the default \textsc{qsogen} dust law and a two-component SED model which includes a scattered light component are presented by the orange lines and triangles. In both cases, a two-component model with the default \textsc{qsogen} dust extinction law is favoured.}
 \label{fig:Orion_dust_law}
\end{figure}

As an alternative to the dust-reddened quasar + scattered light SED model, we attempt to reproduce the UV excess using a single-component dust-reddened quasar SED model, which utilised the Orion dust-extinction law (Section \ref{sec:orion}). We find that not a single HRQ favours the Orion dust extinction law over the dust-reddened quasar + scattered light SED model based on their $\overline{\chi}^2_\nu$ statistics. As an example, in Fig. \ref{fig:Orion_dust_law} we present the best-fit SED models for VIKJ2115-5913 and ULASJ0144-0114 comparing the Orion extinction versus the two-component SED fits. These two examples are representative of the full range in UV/optical continuum shapes in the HRQ sample. The reduced chi squared statistics for VIKJ2115-5913 (ULASJ0144-0114) are $\overline{\chi}^2_\nu=1.3\, (1.1)$ and $\overline{\chi}^2_\nu=3.9\, (18.5)$ for the two-component and Orion dust-law SED models, respectively.

The Orion dust extinction law \citep[Figure 1;][]{2025ApJ...980...36L} is much flatter than that of \textsc{qsogen} \citep[Figure 11;][]{2021MNRAS.508..737T}. Hence, the Orion dust extinction law can broadly reproduce the shape of the HRQ continua at rest-UV wavelengths without the need for a secondary scattered component. However, because the Orion dust law is inherently flat, it is unable to produce sufficiently red rest-optical continua to successfully fit HRQ SEDs and therefore our two-component SED model is always preferred. 

\section{The impact of infrared selection on hot dust amplitudes} \label{App:IR_Selection}

\begin{figure}
\centering
\begin{tabular}{c}
 \includegraphics[scale=0.49, trim={0.0cm, 1.26cm, 0.0cm, 0.0cm}, clip]{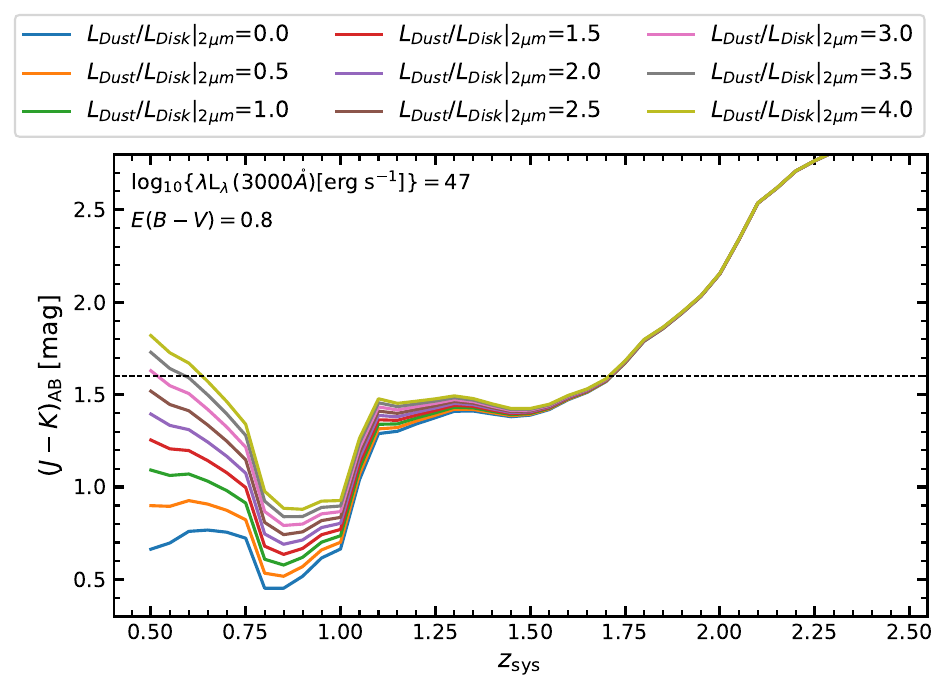}  \\
 \includegraphics[scale=0.49, trim={0.0cm, 0.2cm, 0.0cm, 2.45cm}, clip]{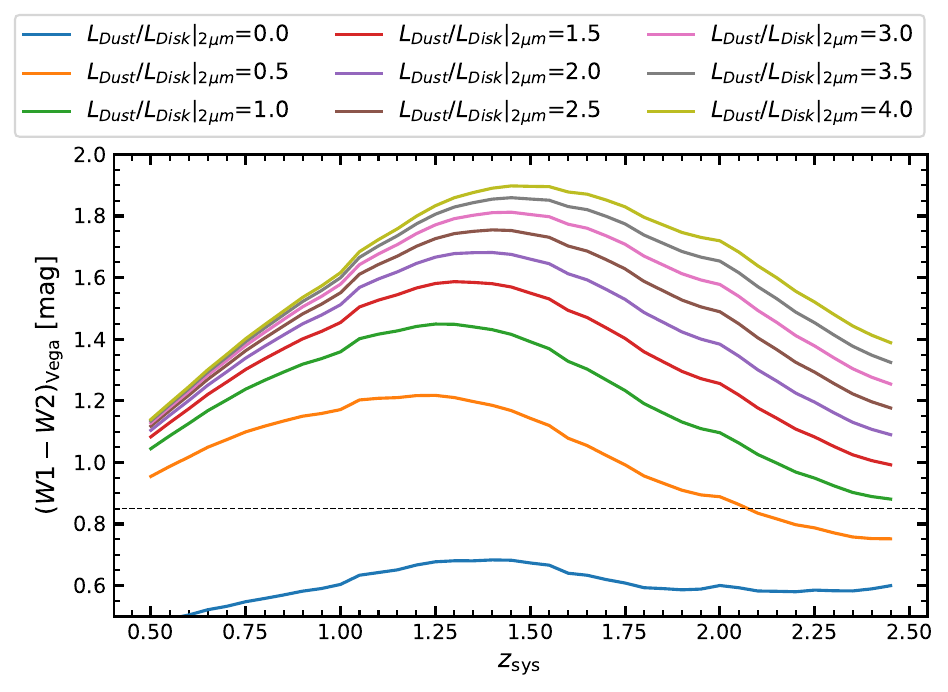} \\
\end{tabular}

  \caption{We present the \mbox{$(J-K)_{\rm AB}$} (top) and \mbox{$(W1-W2)$} (bottom) colours of various \textsc{qsogen} SED models as a function of the systemic redshift. Our \mbox{$(J-K)_{\rm AB} =1.6$} mag (top) and  \mbox{$(W1-W2) = 0.85$} (bottom) colour-selections are illustrated by the black dashed lines. At $z_{\rm sys}\geq1.5$, the \mbox{$(J-K)_{\rm AB}$} colour is insensitive to the hot dust amplitude. Below this redshift, the distribution is shaped by selection effects. The \mbox{$(W1-W2)$} colour-selection appears to bias our sample towards HRQs with higher hot dust amplitudes. For both figures, we assume an extinction $\rm E(B-V)$ = 0.8 and a 3000\AA\, continuum luminosity log$_{10}\{\lambda$L$_{\lambda} (3000\mathring{A})\} = 47.0\; \rm erg\;s^{-1}$ - consistent with the mean extinction and luminosity observed in the 63 HRQs for which an SED model was conducted.}
 \label{fig:NIR_Selection}
\end{figure}

To understand whether the result presented in Section \ref{sec:dust} is biased by the infrared selection of the HRQ sample, we use \textsc{qsogen} to generate a family of models with various sublimation-temperature dust amplitudes and trace how the \mbox{$(J-K)_{\rm AB}$} and \mbox{$(W1-W2)$} colour evolves with redshift. In all instances we assume a line-of-sight dust extinction $\rm E(B-V) = 0.8$ and a 3000\AA\, continuum luminosity log$_{10}\{\lambda$L$_{\lambda} (3000\mathring{A}) [\rm erg\;s^{-1}]\} = 47.0$, which represent the sample averages. The results are presented in Fig. \ref{fig:NIR_Selection}. 

\autoref{fig:NIR_Selection} (top) illustrates how the \mbox{$(J-K)_{\rm AB}$} colour is insensitive to the hot dust at redshifts $z_{\rm sys}\geq1.5$. When comparing hot dust amplitudes across populations, we consider only HRQs with $z_{\rm sys}\geq1.5$. This is because the hot dust amplitude in the lower redshift subsample is largely shaped by near-infrared selection effects - i.e. only the highest dust amplitudes produce the required \mbox{$(J-K)_{\rm AB}$} colour at low redshifts. Conversely, Fig. \ref{fig:NIR_Selection} (bottom) illustrates the strong dependence of the \mbox{$(W1-W2)$} colour on the hot dust amplitude. At $z_{\rm sys}\gtrsim2.0$, HRQs with hot dust amplitudes $\rm L_{Dust}/L_{Disk}|_{2\mu m}<1.0$ may be excluded from the sample. We note that increasing the line-of-sight extinction $\rm E(B-V)$ enables HRQs with lower hot dust amplitudes to meet the \mbox{$(W1-W2)$} colour-selection.

\bsp	
\label{lastpage}

\end{document}